\documentclass[10pt,a4paper]{article} 
\usepackage{graphicx}
\usepackage{amsfonts,amsmath,amssymb}
\usepackage{hyperref}
\usepackage{cite}
\newcommand{\bse}{\begin{subequations}}
\newcommand{\ese}{\end{subequations}}

\begin{document}
\title{\bf Thermodynamic scaling and string dynamics in rotating holographic QCD}

\date{}
\maketitle
\vspace*{-0.3cm}
\begin{center}
{\bf Leila Shahkarami$^{1}$, Farid Charmchi$^{2}$}\\
\vspace*{0.3cm}
{\it {School of Physics, Damghan University, Damghan 36716-45667, Iran
}} \\
\vspace*{0.3cm}
{\it  {${}^1$l.shahkarami@du.ac.ir}, {${}^2$farid.charmchi@gmail.com} }
\end{center}

\begin{abstract}
We investigate the effect of finite angular velocity on the phase structure and Schwinger pair production in a holographic QCD model based on the Einstein--Maxwell--dilaton framework. Rotation is introduced through a boost construction that generates a stationary rotating geometry from the corresponding static background. We first study the thermodynamic phase diagram using the grand potential together with several response functions and show that all thermodynamic phase boundaries satisfy an exact scaling relation under rotation, implying that the rotating thermodynamic sector is completely determined by the corresponding static solution through a simple boost transformation. We then investigate confinement through the effective string tension and find that, consistent with the thermodynamic observables, rotation decreases the string confinement--deconfinement transition temperature and chemical potential. The corresponding phase boundary, however, no longer obeys the thermodynamic scaling relation. Consequently, the region in the phase diagram that is thermodynamically confined but string deconfined grows with angular velocity. We further study the Schwinger effect using the potential analysis approach. Rotation lowers both the confining and catastrophic critical electric fields and suppresses the height and width of the potential barrier, thereby enhancing pair production in both confined and deconfined phases. In the deconfined phase, rotation also generates a worldsheet horizon that limits the radial extent of connected string solutions and reduces both the maximum quark--antiquark separation and the deepest turning point of the energetically favored string. These results demonstrate that, while rotation acts trivially in the thermodynamic sector through an exact scaling law, it produces genuinely new effects in the string sector.
\end{abstract}
\section{Introduction}
The quark--gluon plasma (QGP) produced in noncentral heavy-ion collisions is expected to possess not only extremely high temperatures and energy densities, but also intense electromagnetic fields and enormous angular momentum. While the magnetic field generated during the early stages of the collision can reach $eB\sim10^{18}$--$10^{20}$ Gauss \cite{Bmagnitude1,Bmagnitude2}, the total angular momentum may be as large as $\mathcal{O}(10^4\!-\!10^5)\hbar$ \cite{angularm}, with estimated average angular velocities of order $\omega\sim  10^{22}~{\rm s}^{-1}$ \cite{star2017} corresponding to tens of ${\rm MeV}$ in natural units, while hydrodynamic simulations suggest values as large as $\omega \sim 20$--$40 ~{\rm MeV}$ \cite{hydro}. The observation of global polarization of $\Lambda$ and $\bar{\Lambda}$ hyperons by the STAR Collaboration \cite{star2017} and the subsequent measurements of vector-meson spin alignment by the ALICE and STAR Collaborations \cite{alice,star2023} have provided compelling experimental evidence that rotational effects play an important role in strongly interacting matter. 
Rotation is expected to give rise to a variety of novel phenomena, including the chiral vortical effect \cite{chiralvortical1,chiralvortical2,chiralvortical3}, spin polarization \cite{polarization1,polarization2,polarization3}, and modifications of the chiral restoration and deconfinement phase transitions \cite{transition1,huang,transition2,transition3,transition4,transition5,effective1,effective2,effective3,effective4,effective6,
effective7,effective8,effective9,effective10,lattice1,lattice2,lattice3,correctphase1,correctphase2}. These developments have stimulated extensive theoretical studies of rotating QCD using effective field theories, lattice simulations and holographic approaches.

One of the central questions concerns the influence of rotation on the QCD phase structure. Most of the studies in effective QCD models, such as the Nambu--Jona-Lasinio and hadron resonance gas models, generally predict that increasing the angular velocity lowers the critical temperature of the chiral and/or deconfinement phase transitions \cite{effective1,effective2,effective3,effective4,effective6,effective7,effective8,effective9,effective10}. 
In contrast, present lattice simulations indicate an enhancement of both the chiral restoration and deconfinement transition temperatures with increasing angular velocity \cite{lattice1,lattice2,lattice3}. 
This qualitative discrepancy has stimulated considerable efforts to clarify the role of gluodynamics in rotating matter and to construct theoretical descriptions capable of reproducing the lattice results \cite{correctphase1,correctphase2}. Although important progress has recently been achieved, the influence of rotation on strongly interacting matter remains an active subject of investigation, making it worthwhile to explore the problem from different nonperturbative perspectives and through different physical observables.

The gauge/gravity duality provides an alternative nonperturbative framework for investigating strongly coupled gauge theories \cite{adscft1,adscft2,adscft3,adscft4}. Within this approach, several different realizations of rotation have been proposed, each designed to capture different aspects of rotating QCD matter. Exact Kerr--AdS black-hole geometries provide genuine rotating gravitational solutions but are dual to conformal field theories and are therefore mainly applicable to high-temperature plasmas \cite{KerrAds1,KerrAds2,KerrAds3,KerrAds4,KerrAds5,KerrAds6}. A widely adopted alternative introduces rotation through local Lorentz transformations or boost constructions acting on known static holographic backgrounds \cite{huang,boost1,boost2,boost3,boost4,boost5,boost6}. This approach preserves an exact analytic relation between the rotating and static geometries, making it possible to distinguish effects that arise purely from the boost from genuinely new rotational phenomena.
Another realization starts from holographic backgrounds written in cylindrical coordinates and introduces rotation through a global transformation of the angular coordinate,
$\theta\rightarrow\theta+\omega t$ \cite{planar}. Since the resulting geometry preserves the cylindrical structure, it naturally distinguishes directions parallel and perpendicular to the rotation axis and is therefore particularly suitable for investigating rotational anisotropies in quark--antiquark observables. Because the gravitational background itself is not modified, however, the thermodynamic properties remain those of the original static solution, while rotation manifests itself only through the dynamics of the probe sector.
Rotation has also been incorporated through rotating boundary gauge fields in probe-limit holographic models, enabling studies of chiral symmetry breaking while neglecting the backreaction of the rotating matter on the bulk geometry \cite{rotatinggauge1,rotatinggauge2,rotatinggauge3,rotatinggauge4}. More recently, a fully backreacted anisotropic Einstein--Maxwell--dilaton (EMD) background sourced by rotational gauge fields has been constructed \cite{correctphase2}. This background successfully reproduces the enhancement of the transition temperature observed in present lattice simulations. The complexity of the coupled Einstein equations, however, requires working within a near-axis approximation, where the geometry is assumed to depend only on the holographic coordinate. Consequently, the construction is intended to describe the vicinity of the rotation axis for phenomenologically relevant angular velocities rather than the full rotating plasma.
Rather than competing with one another, these constructions provide complementary descriptions of rotating strongly coupled matter, each with its own advantages and range of applicability.

This qualitative discrepancy has stimulated considerable efforts to clarify the role of gluodynamics in rotating matter and to construct theoretical descriptions capable of reproducing the lattice results [33, 34]. In particular, fully back-reacted anisotropic EMD geometries can reverse the sign of the shift and raise the transition temperature, in agreement with lattice data. The boost construction employed in the present work, by contrast, inherits the model-like lowering of the critical temperature (as we confirm below). Although important progress has recently been achieved, the influence of rotation on strongly interacting matter therefore remains an active subject of investigation, making it worthwhile to explore the problem from different nonperturbative perspectives and through different physical observables.

In the present work we employ the second construction based on a local Lorentz transformation of the static EMD background developed in \cite{dudalm,huang}. This framework preserves an exact analytic relation between the rotating and static geometries while allowing quark--antiquark observables to be investigated at arbitrary distances $l$ from the rotation axis within the physical region $\omega l<1$. It therefore provides a unique setting in which effects determined entirely by the boost can be disentangled from genuinely new rotational phenomena associated with string dynamics.

Among the various string observables, the Schwinger effect \cite{Schwinger} provides a particularly sensitive probe of the interplay between confinement and external electromagnetic fields. Within the holographic approach, it has been extensively investigated through the potential analysis method and the Dirac--Born--Infeld action in a wide variety of strongly coupled systems \cite{semenoff,potential,Sch1,Sch2,Sch3,confin1,confin2,Sch4,Sch5,confinrev,dehghani,magneticdecay,LF,us2,me,Sch6,Sch7,Sch8}. Existing holographic studies of the Schwinger effect in rotating backgrounds have mainly focused on demonstrating that rotation enhances pair production by reducing the potential barrier and the critical electric field \cite{rotSch1,rotSch2,rotSch3}. While these analyses establish that rotation facilitates vacuum decay, they do not address how rotation modifies the confinement mechanism itself, the confining critical electric field associated with the infrared wall, or the relation between thermodynamic and string-sector probes. A comprehensive investigation of the Schwinger effect in rotating holographic models with confinement therefore remains lacking.

The static EMD model adopted here \cite{dudalm} possesses two additional features that make it particularly suitable for the present study. First, unlike many holographic QCD models in which the confined phase is described by a temperature-independent thermal-gas solution, both the confined and deconfined phases are described by black-hole geometries whose properties depend continuously on the temperature and chemical potential. After introducing rotation, the background therefore depends simultaneously on $(T,\mu,\omega)$ throughout the entire phase diagram, making it possible to investigate the rotational response of string observables associated with finite-mass quark--antiquark pairs in both phases within a unified framework.
Second, the model exhibits a rich phase structure in the $(T,\mu)$ plane, consisting of a first-order confinement--deconfinement transition terminating at a critical end point (CEP), followed by a supercritical region. A detailed analysis of both thermodynamic and string observables in the static background was presented in \cite{superus}. Besides the thermodynamic transition lines, that study identified an additional confinement criterion associated with the existence of a dynamical infrared wall. Unlike the thermodynamic phase boundaries, the corresponding string confinement line does not originate from the CEP, giving rise to a region where the system remains thermodynamically confined while string observables already exhibit deconfined behavior.
The boost construction provides a setting in which the rotating geometry is known analytically from the corresponding static solution, which simultaneously modifies the thermodynamic and string sectors in the presence of the rotation. This makes it possible to compare these observables on the same rotating background. This construction also provides a rare opportunity to distinguish rotational effects that are purely kinematical consequences of the boost from genuinely dynamical effects associated with different physical probes. 

In order to address these questions, we first construct the thermodynamic phase diagrams of the rotating plasma from the grand potential and several response functions and compare them with the corresponding static results. We then investigate confinement from the viewpoint of fundamental strings through the effective string tension and determine the associated confinement--deconfinement phase boundaries. Finally, we study the Schwinger effect by analyzing the critical electric fields, the quark--antiquark separation length and the total potential in both confined and deconfined phases. Together, these analyses provide a comprehensive picture of how rotation influences both the thermodynamic and string sectors of the holographic plasma.

The remainder of this paper is organized as follows. In Sec.\,2 we briefly review the rotating Einstein--Maxwell--dilaton background and its thermodynamic properties. Section 3 is devoted to the thermodynamic phase structure. In Sec.\,4 we investigate confinement and the Schwinger effect from the viewpoint of the string sector. Finally, Sec.\,5 contains our summary and discussion.

\section{Holographic rotating Einstein--Maxwell--Dilaton background}

In this section, we introduce the gravitational background employed throughout this work. We first review the static Einstein--Maxwell--Dilaton (EMD) geometry describing the dual plasma at finite temperature and chemical potential \cite{dudalm}, and then construct its rotating extension following the approach proposed in \cite{huang}.

\subsection{Static EMD background}

We consider the five-dimensional Einstein--Maxwell--Dilaton action
\begin{align}
S=\frac{1}{16\pi G_5}\int d^5x\,\sqrt{-g}\,
\left[
R-\frac{1}{2}(\partial\phi)^2-V(\phi)-\frac{f(\phi)}{4}F_{\mu\nu}F^{\mu\nu}
\right],
\label{action}
\end{align}
where \(R\) denotes the Ricci scalar, \(\phi\) is the dilaton field, \(V(\phi)\) is the dilaton potential, and \(f(\phi)\) is the gauge kinetic function associated with the Maxwell field \(A_\mu\), with \(F=dA\). Finite baryon density is introduced through a nonvanishing temporal component of the gauge field.

For the static plasma background, we employ the black-hole ansatz
\begin{align}
ds^2=\frac{e^{2A_e(z)}}{z^2}
\left[
-G(z)\,dt^2+dx_1^2+dx_2^2+dx_3^2+\frac{dz^2}{G(z)}
\right],
\qquad
A=A_t(z)\,dt,
\qquad
\phi=\phi(z),
\label{staticmetric}
\end{align}
where the holographic coordinate \(z\) runs from the AdS boundary at \(z=0\) to the black-hole horizon located at \(z=z_h\), determined by
\begin{align}
G(z_h)=0.
\label{horizon}
\end{align}
The warp factor \(A_e(z)\) characterizes the deviation from conformality, while the blackening function \(G(z)\) determines the thermal properties of the system. The thermodynamic states of the dual theory are specified by the horizon position \(z_h\) and the chemical-potential parameter \(\mu_0\), defined through the boundary value of the temporal gauge field,
\begin{align}
\mu_0=\lim_{z\to0}A_t(z)-A_t(z_h).
\label{mu0def}
\end{align}
Imposing the regularity condition \(A_t(z_h)=0\), one simply obtains \(\mu_0=A_t(0)\). Throughout this work, the AdS radius is set to unity.

The equations of motion are obtained by varying the action with respect to the metric \(g_{\mu\nu}\), the dilaton field \(\phi\), and the gauge field \(A_\mu\). Their explicit form can be found in \cite{dudalm} and will not be repeated here.
Physical observables of the boundary theory can then be extracted using the holographic dictionary by solving the background equations of motion. In bottom-up holographic QCD models, the functions \(V(\phi)\) and \(f(\phi)\) are usually chosen in such a way as to reproduce the desired infrared behavior and thermodynamic properties of QCD. In the present work, however, following \cite{dudalm}, we employ the potential reconstruction approach \cite{reconst1,reconst2,reconst3,reconst4,reconst5,hajilou}.
Within this framework, the deformation function \(A_e(z)\) and the gauge kinetic function \(f(\phi)\) are chosen as
\begin{align}
A_e(z)= -\frac{3}{4}\ln(az^2+1)
+\frac{1}{2}\ln(bz^3+1)
-\frac{3}{4}\ln(dz^4+1),
\label{warp}
\end{align}
and
\begin{align}
f(\phi(z))=e^{-cz^2-A_e(z)},
\label{gaugekinetic}
\end{align}
which were shown to provide a good description of QCD thermodynamics.

Once these functions are specified, the dilaton potential \(V(\phi)\), the temporal gauge field \(A_t(z)\), and the blackening function \(G(z)\) can be determined from the equations of motion. Within the potential reconstruction approach, the reconstructed dilaton potential inherits an implicit dependence on the thermodynamic parameters through the background solution. The resulting expressions for \(A_t(z)\) and \(G(z)\) are given by
\begin{align}
A_t(z)=\mu_0\,\frac{e^{-cz^2}-e^{-cz_h^2}}{1-e^{-cz_h^2}},
\label{At}
\end{align}
and
\begin{align}
G(z)&=
1-\frac{\displaystyle \int_0^z dx\,x^3e^{-3A_e(x)}}
{\displaystyle \int_0^{z_h}dx\,x^3e^{-3A_e(x)}}+\frac{2c\mu_0^2}
{\left(1-e^{-cz_h^2}\right)^2}
\nonumber\\
&\quad
\times
\frac{
\displaystyle
\int_0^{z_h} dx\,x^3e^{-3A_e(x)}
\int_{z_h}^z dx\,x^3e^{-3A_e(x)-cx^2}
-
\displaystyle
\int_0^{z_h} dx\,x^3e^{-3A_e(x)-cx^2}
\int_{z_h}^z dx\,x^3 e^{-3A_e(x)}
}
{\displaystyle \int_0^{z_h}dx\,x^3e^{-3A_e(x)}}.
\label{blackening}
\end{align}

A complete solution of the system is characterized by the parameters \(z_h\), \(\mu_0\), \(a\), \(b\), \(c\), and \(d\). The parameters \(a\), \(b\), \(c\), and \(d\) are fixed by requiring the model to reproduce lattice results for the pure gluon system at zero chemical potential, leading to \cite{dudalm}
\begin{align}
a&=0.1289~\text{GeV}^2,
\nonumber\\
b&=0.3625~\text{GeV}^3,
\nonumber\\
c&=1.16~\text{GeV}^2,
\nonumber\\
d&=0.1289~\text{GeV}^4.
\label{parameters}
\end{align}

Having established the static EMD background, we now turn to the construction of its rotating extension.

\subsection{Rotating geometry}

A fully dynamical description of a rotating strongly coupled plasma would require solving the Einstein equations for a genuinely rotating metric ansatz subject to appropriate boundary conditions. Exact rotating solutions, however, are generally restricted to special cases such as Kerr-AdS, BTZ, or related geometries. For phenomenological bottom-up holographic QCD models such as the EMD system considered here, obtaining analytic rotating solutions is considerably more difficult and typically requires extensive numerical analysis.

Instead, following \cite{huang}, we construct a stationary rotating extension of the EMD background by compactifying one spatial direction and performing a Lorentz boost in the compactified direction. 
To implement this construction, one first compactifies one of the spatial directions, say \(x_3\), according to
\begin{align}
x_3=l\,\theta,
\label{compactification}
\end{align}
where \(\theta\) is the compact angular coordinate with periodicity \(2\pi\), and \(l\) denotes the radius of the compact circle. The spatial manifold of the boundary theory therefore acquires the topology $\mathbb{R}^2\times S^1$.
Rotation is then introduced through the Lorentz transformation
\begin{align}
t \rightarrow \gamma\,(t+\omega l^2\theta), \quad 
\theta \rightarrow \gamma\,(\theta+\omega t),
\label{boost}
\end{align}
where \(\omega\) denotes the angular velocity and
\begin{align}
\gamma=\frac{1}{\sqrt{1-\omega^2 l^2}}
\label{gamma}
\end{align}
is the corresponding Lorentz factor. This transformation mixes the temporal and angular coordinates and generates an off-diagonal metric component, thereby converting the static background into a stationary rotating geometry.

In a fully rotating system, the tangential velocity depends on the radial distance from the rotation axis, making the geometry intrinsically position dependent. To retain analytic control over the system, we restrict the analysis to a fixed radial distance \(l\) from the rotation axis. The local tangential velocity is then given by
\begin{align}
v=\omega l.
\label{velocity}
\end{align}
As long as \(v<1\), the rotating geometry remains physically admissible and the Killing vector \(\partial_t\) remains timelike outside the horizon.

The resulting rotating metric takes the form
\begin{align}
ds^2=H(z)\left[
-N(z)\,dt^2
+\frac{\,dz^2}{G(z)}
+R(z)\bigl(d\theta+P(z)\,dt\bigr)^2
+dx_1^2+dx_2^2\right],
\label{rotatingmetric}
\end{align}
where
\begin{align}
N(z)&=
\frac{G(z)\left(1-\omega^2 l^2\right)}
{1-G(z)\omega^2 l^2},
\label{Nz}\\
R(z)&=
\gamma^2 l^2
-G(z)\gamma^2\omega^2 l^4,
\label{Rz}\\
P(z)&=
\frac{\omega-G(z)\omega}
{1-G(z)\omega^2 l^2},
\label{Pz}
\end{align}
and
\begin{align}
H(z)=\frac{e^{2A_e(z)}}{z^2}.
\label{Hz}
\end{align}
The off-diagonal component proportional to \(P(z)\) reflects the stationary nature of the boost-generated rotating geometry. The transverse directions \(x_1\) and \(x_2\) remain equivalent, indicating that the present setup probes the rotation-symmetric sector of the plasma.

In the rotating background, the chemical potential is no longer determined solely by the temporal component of the gauge field. Since the horizon is generated by the Killing vector
\begin{align}
\chi=\partial_t-\omega\,\partial_\theta,
\label{killing}
\end{align}
the gauge-invariant chemical potential is obtained from the contraction \(A_\mu\chi^\mu\). Accordingly, the chemical potential is defined as the potential difference between the boundary and the horizon,
\begin{align}
\mu=
\left.A_\mu\chi^\mu\right|_{z=0}
-
\left.A_\mu\chi^\mu\right|_{z_h}.
\label{chemicalpotential}
\end{align}
Starting from the static configuration characterized by the parameter \(\mu_0\), the rotating geometry is generated through the Lorentz transformation in Eq.\,\eqref{boost}. Under this transformation, the gauge field acquires an angular component proportional to \(\omega A_t\) and becomes \(A_{\mu}=\gamma A_t \delta_{\mu}^t+\omega\gamma A_t \delta_{\mu}^{\theta}\). Using Eq.\,\eqref{chemicalpotential}, one obtains
\begin{align}
\mu
=
\mu_0\sqrt{1-\omega^2 l^2}
=
\frac{\mu_0}{\gamma},
\label{murot}
\end{align}
or equivalently,
\begin{align}
\mu_0=\gamma\mu.
\label{mu0rot}
\end{align}

Similarly, within this setup, the physical temperature of the rotating plasma is obtained from the surface gravity associated with the horizon-generating Killing vector in Eq.\,\eqref{killing}. The resulting temperature is related to the temperature parameter of the static background through
\begin{align}
T_0=\gamma T.
\label{temperatureboost}
\end{align}
This relation also follows directly from the explicit evaluation of the Hawking temperature in the rotating geometry, which will be discussed in the next section. Therefore, the effective thermodynamic parameters measured in the rotating frame are modified by the boost-induced Lorentz factor, and the static variables \((T_0,\mu_0)\) are rewritten in terms of the physical parameters \((T,\mu,\omega)\) of the rotating plasma.

It should be emphasized that the present construction does not describe a fully dynamical rotating geometry. Rather, it provides a boost-generated stationary spacetime restricted to a fixed-\(l\) slice. From the viewpoint of the dual field theory, the setup should therefore be interpreted as an effective local description of a rotating plasma. Although the geometry is generated kinematically through a boost transformation, the resulting Lorentz factors modify the local thermodynamic quantities as well as the phase structure and may affect the thermodynamic and nonperturbative observables such as vacuum pair production in qualitatively different ways.

In the following sections, we employ this rotating EMD background to investigate the influence of rotation on the thermodynamics of the system, the location of the critical end point, and the behavior of the Schwinger effect across the first-order, second-order, and supercritical regions of the phase diagram.

\section{Thermodynamics of the rotating plasma}

Having established the rotating EMD background, we now investigate the thermodynamic properties of the dual plasma and their response to the angular velocity. In the present construction, the rotating geometry is generated from the static solution through the boost transformation discussed in the previous section. As a consequence, the physical thermodynamic parameters $(T,\mu)$ are related to the corresponding quantities of the static background through the Lorentz factor $\gamma$.

An important question is therefore whether the resulting thermodynamic behavior exhibits genuinely new features induced by rotation, or whether it is entirely determined by the boost-induced mapping between the rotating and static configurations. To address this issue, we first study the basic thermodynamic observables of the system and then analyze the phase structure using several independent probes, including the specific heat, baryon number susceptibility, and the speed of sound. This allows us to determine the confinement--deconfinement transition line, the location of the critical end point, and the crossover region in the $(T,\mu,\omega)$ parameter space.

\subsection{Thermodynamic quantities}

We begin by introducing basic thermodynamic observables employed throughout this work. Since the dual plasma is studied at finite temperature and chemical potential, the appropriate framework is the grand canonical ensemble, in which the thermodynamic state is specified by the variables $(T,\mu,\omega)$. The physical temperature and chemical potential are related to the corresponding parameters of the static background through Eqs.\,\eqref{mu0rot} and \eqref{temperatureboost}. Consequently, all thermodynamic quantities may be expressed as functions of $(T,\mu,\omega)$.

The temperature of the rotating plasma is determined by the surface gravity associated with the horizon-generating Killing vector \eqref{killing}. Denoting the corresponding surface gravity by $\kappa$, the Hawking temperature is given by
\begin{align}
T=&\left|\frac{\kappa}{2\pi}\right|\nonumber\\
=&\frac{z_h^3\!~e^{-3A_e(z_h)}}{\gamma}\,
\frac{\left(1-e^{-cz_h^2}\right)^2+2c\mu_0^2\left(e^{-cz_h^2}\int^{z_h}_0\! dx\!~x^3\!~e^{-3A_e(x)}
-\int^{z_h}_0\! dx\!~x^3\!~e^{-3A_e(x)}e^{-cx^2}\right)}{4\pi \left(1-e^{-cz_h^2}\right)^2 \int^{z_h}_0\! dx\!~x^3\!~e^{-3A_e(x)}}
,
\label{temperature}
\end{align}
where the explicit expression in the second equality follows from the rotating geometry \eqref{rotatingmetric}.
The entropy density is obtained from the Bekenstein--Hawking relation,
\begin{align}
s=
\frac{\gamma \,l}{4G_5}
\frac{e^{3A_e(z_h)}}{z_h^3}.
\label{entropy}
\end{align}
 Throughout this work, we fix the five-dimensional Newton constant according to $G_5=\frac{45\pi}{16N_c^2}$, 
which ensures that the Stefan--Boltzmann law $p=\frac{\pi^2}{45}N_c^2T^4$
is reproduced in the ultraviolet regime of asymptotically large temperatures, in the static case with $\omega=0$. In the following, all thermodynamic quantities are normalized by $N_c^2$.

The remaining thermodynamic observables are obtained from the grand potential density $\Omega$. In the present boost-generated construction, we treat $\omega$ as an external parameter characterizing the geometry rather than as an independent thermodynamic variable with a known holographic conjugate charge. Consequently, the thermodynamic relations employed throughout this work involve variations with respect to $T$ and $\mu$ at fixed $\omega$. 
The grand potential is evaluated for fixed values of the angular velocity.  Its differential form is
\begin{align}
d\Omega=-s \,dT-\rho\, d\mu,
\label{dOmega}
\end{align}
where $\rho$ denotes the baryon number density.
At fixed values of $\mu$ and $\omega$, the grand potential can be written as
\begin{align}
\Omega=-\int s\, dT,
\label{grandpotential}
\end{align}
up to an integration constant which is fixed by requiring $\Omega(z_h\rightarrow \infty)=0$. This choice corresponds to taking the thermal-gas background as the reference state, which coincides with the black-hole solution in the limit $z_h\rightarrow\infty$ at vanishing chemical potential.

The pressure is then given by
\begin{align}
p=-\Omega,
\label{pressure}
\end{align}
while the energy density follows from the first law of thermodynamics,
\begin{align}
\epsilon=-p+Ts+\mu\rho.
\label{energydensity}
\end{align}
The trace anomaly,
\begin{align}
I=\epsilon-3p,
\label{traceanomaly}
\end{align}
provides a measure of the deviation from conformality and plays an important role in characterizing the strongly coupled plasma near the transition region.
The baryon number density moreover is defined thermodynamically through
\begin{align}
\rho
=-\left(
\frac{\partial\Omega}{\partial\mu}
\right)_T .
\label{density}
\end{align}
In practice, $\rho$ is obtained holographically from the near-boundary behavior of the temporal component of the gauge field. Expanding $A_t(z)$ close to the AdS boundary,
\begin{align}
A_t(z)=\mu-\rho\,z^2+\mathcal{O}(z^4),
\end{align}
one finds
\begin{align}
\rho
=\frac{c\,\mu_0}{e^{cz_h^2}-1},
\label{rhobulk}
\end{align}
where the relation between $\mu_0$ and the physical chemical potential is given by Eq.~\eqref{murot}.

\begin{figure}[h]
\begin{center}
\includegraphics[width=6.8cm]{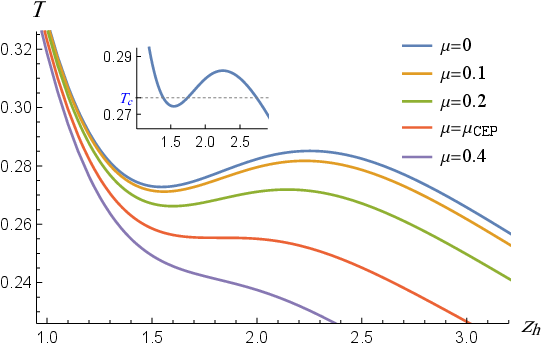}\hspace{.3cm}
\includegraphics[width=6.8cm]{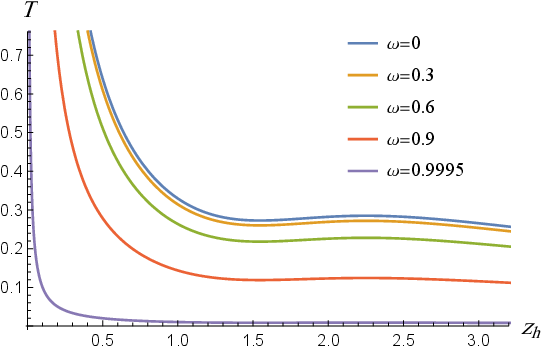}
\end{center}
\caption{\footnotesize 
Left (right) graph shows the temperature as a function of the horizon radius $z_h$ at $\omega=0$ ($\mu=0$) for various values of the chemical potential (angular velocity).}
\label{Tzh}
\end{figure} 

Having established the thermodynamic framework, we now investigate the behavior of these quantities in the presence of rotation.
In what follows, to simplify the presentation, we omit the units. All the numbers given in the figures are in units $\mathrm{GeV}$.

\begin{figure}[th]
\begin{center}
\includegraphics[width=6.8cm]{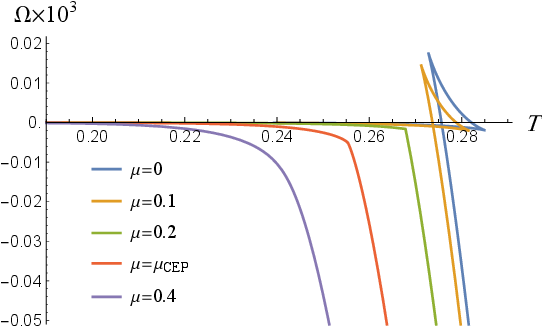}\hspace{.3cm}
\includegraphics[width=6.8cm]{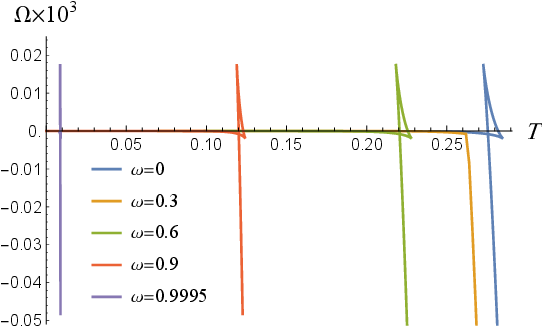}
\end{center}
\caption{\footnotesize 
Left (right) graph shows the grand potential as a function of temperature at $\omega=0$ ($\mu=0$) for various values of the chemical potential (angular velocity).}
\label{FT}
\end{figure} 

The thermodynamic behavior of the system is most conveniently understood by first examining the temperature as a function of the horizon position. Figure \ref{Tzh} displays $T(z_h)$ for several values of the chemical potential at vanishing angular velocity, while the right graph of Fig.\,\ref{Tzh} shows the corresponding curves for different angular velocities at $\mu=0$. For sufficiently small chemical potentials, the temperature exhibits a characteristic nonmonotonic behavior and develops two extrema. As a consequence, three distinct black-hole branches coexist within a finite temperature interval. 
The branch with small horizon radius $z_h$ and negative slope corresponds to a stable high-temperature solution, while the branch with large horizon radius and negative slope represents a stable low-temperature solution. These two branches are separated by an intermediate branch with positive slope, which is thermodynamically unstable. This structure is the holographic manifestation of the first-order confinement--deconfinement transition and persists up to the critical end point at $\mu_{\mathrm{CEP}}\approx 0.312 \, \mathrm{GeV}$, beyond which the extrema merge and the transition turns into a crossover \cite{superus}. 

\begin{figure}[h]
\begin{center}
\includegraphics[width=6.8cm]{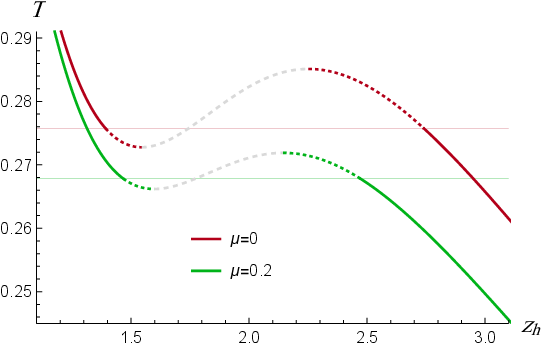}\hspace{.3cm}
\includegraphics[width=6.8cm]{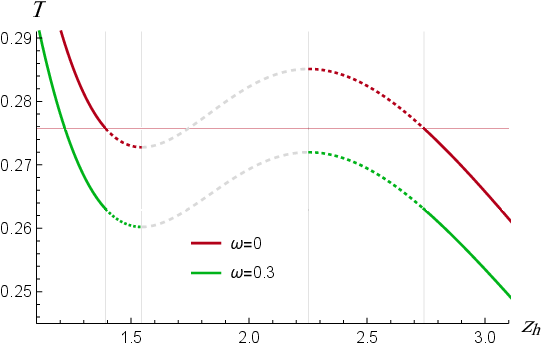}
\end{center}
\caption{\footnotesize 
Stable, metastable and unstable  solutions of $T(z_h)$ for two different values of chemical potential at zero angular velocity (left graph) and for two different values of angular velocity at zero chemical potential (right graph). Horizontal thin lines show the position of the transition temperature of each case, found from the swallow tail structure in the graphs of the grand potential. Moreover, vertical thin lines show the separate branches of the solutions while varying the angular velocity.}
\label{Tzhdetail}
\end{figure} 

The corresponding grand potential diagrams are shown in Fig.\,\ref{FT}. As expected, the coexistence of three black-hole branches gives rise to the familiar swallow-tail structure of $\Omega(T)$, indicating the presence of a first-order phase transition. The transition temperature is determined by the intersection point of the two thermodynamically favored branches, where the grand potential becomes degenerate. As the chemical potential approaches the critical value, the swallow tail gradually shrinks and eventually disappears at the critical end point. 

The effect of the angular velocity is qualitatively different.
Although increasing $\omega$ modifies the physical temperature through the boost factor and shifts the location of the transition, the overall structure of the temperature curves remains essentially unchanged. In particular, the characteristic values of $z_h$ at which the extrema occur are found to be nearly insensitive to the angular velocity, in contrast to the behavior observed when varying the chemical potential. This indicates that the branch structure of the black-hole solutions is primarily governed by the underlying geometry, whereas the rotation introduced through the boost transformation mainly affects the thermodynamic variables through the corresponding Lorentz rescaling. To make the branch structure more transparent, in Fig.\,\ref{Tzhdetail} we separately identify the stable, metastable, and unstable solutions. The thermodynamically preferred branches are represented by solid curves, while the metastable extensions are shown by dashed curves. The unstable branch connecting the two extrema corresponds to the upper segment of the swallow tail in the grand potential and is characterized by a negative heat capacity. 

The resulting picture is completely analogous to that of the static EMD background, with the main effect of the angular velocity being a displacement of the thermodynamic scales rather than a qualitative modification of the black-hole branches themselves.  Whether this observation extends to other thermodynamic observables and to the resulting phase diagram will be examined in the following subsections.

\begin{figure}[h]
\begin{center}
\includegraphics[width=6.8cm]{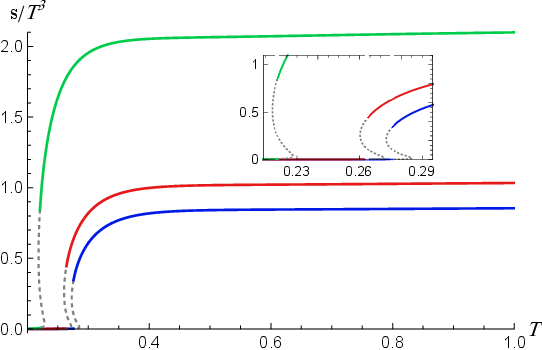}\hspace{.3cm}
\includegraphics[width=6.8cm]{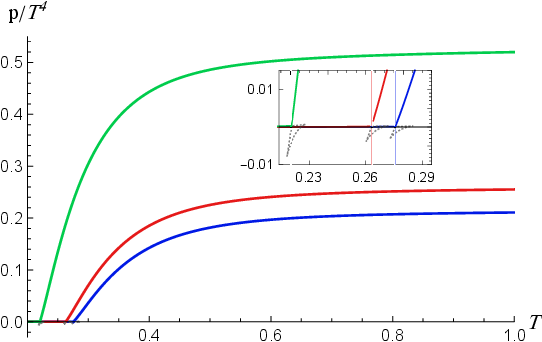}\\\vspace{.2cm}\includegraphics[width=6.8cm]{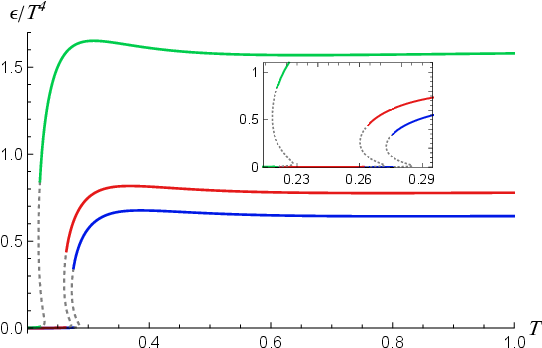}\hspace{.3cm}
\includegraphics[width=6.8cm]{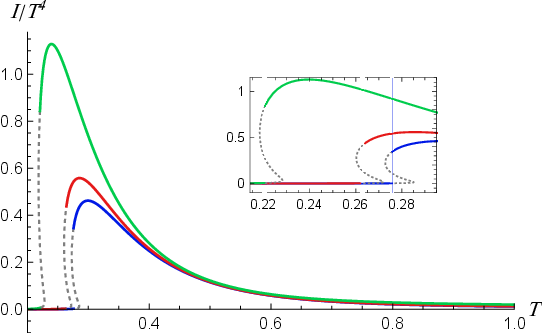}
\end{center}
\caption{\footnotesize 
Thermodynamic properties of the system for various values of angular velocity $\omega=0$ (blue solid line), $\omega=0.3\,\mathrm{GeV}$ (red solid line) and $\omega=0.6\,\mathrm{GeV}$ (green solid line) at vanishing chemical potential. Gray dashed lines show the unstable and metastable branches of the solutions.}
\label{speImu0}
\end{figure}

The behavior of the scaled entropy density, pressure, energy density, and trace anomaly is presented in Fig.\,\ref{speImu0} for several values of the angular velocity at zero chemical potential. In each figure, the stable branches are shown by the solid blue, red, and green curves corresponding to $\omega=0$, $0.3\,\mathrm{GeV}$, and $0.6\,\mathrm{GeV}$, respectively, while the unstable and metastable solutions are represented by the dashed gray curves. The insets display enlarged views of the transition region, where the multivalued structure associated with the first-order phase transition can be seen more clearly.

The first-order transition manifests itself differently in different thermodynamic observables. For the grand potential and the pressure, the coexistence of three black-hole branches gives rise to the familiar swallow-tail structure. In contrast, the entropy density, energy density, and trace anomaly become multivalued within the finite temperature interval bounded by the two extrema of the temperature curve. These features disappear at the critical end point, where the three branches merge into a single solution and the transition becomes second order, as we will see in the next figure.

The effect of the angular velocity on these observables is qualitatively similar for all quantities considered here. Increasing $\omega$ enhances the values of $s/T^3$, $p/T^4$, and $\epsilon/T^4$ over the entire temperature range. The same trend is also observed for the trace anomaly in the vicinity of its peak, although $I/T^4$ approaches zero at sufficiently large temperatures independently of the value of the angular velocity, reflecting the restoration of approximate conformal behavior in the ultraviolet regime. Furthermore, the asymptotic high-temperature values of $s/T^3$, $p/T^4$, and $\epsilon/T^4$ are found to increase with $\omega$ \cite{huang}.

\begin{figure}[h]
\begin{center}
\includegraphics[width=6.8cm]{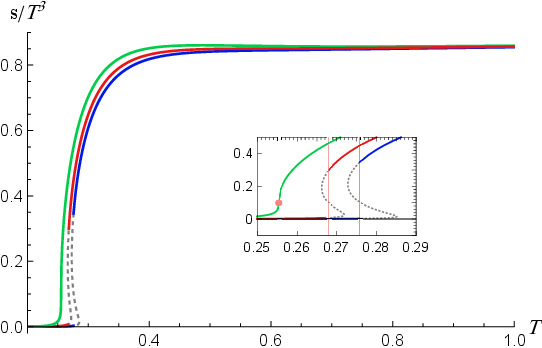}\hspace{.3cm}
\includegraphics[width=6.8cm]{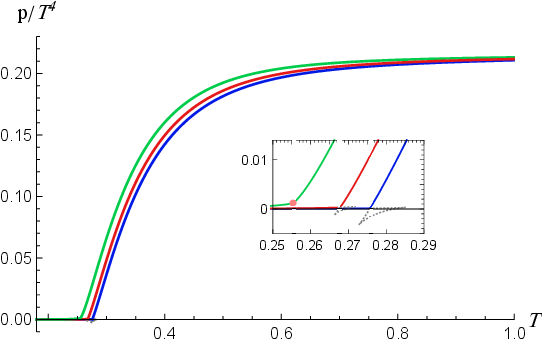}\\\vspace{.2cm}\includegraphics[width=6.8cm]{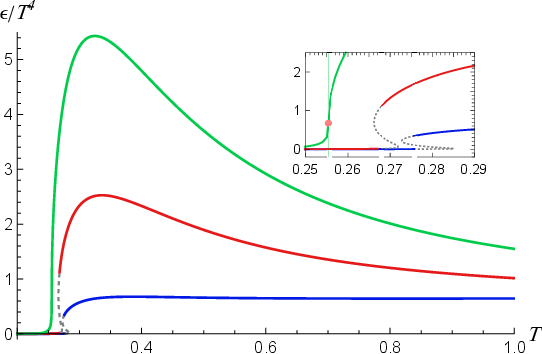}\hspace{.3cm}
\includegraphics[width=6.8cm]{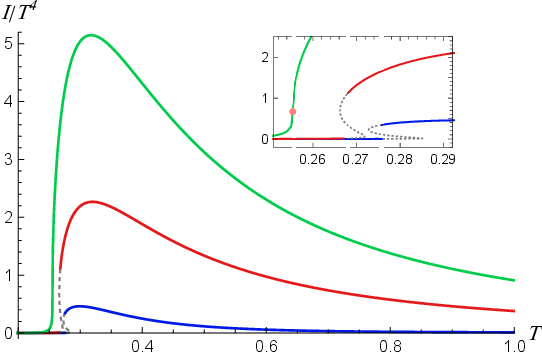}
\end{center}
\caption{\footnotesize 
Thermodynamic properties of the system for various values of chemical potential $\mu=0$ (blue solid line), $\mu=0.2\,\mathrm{GeV}$ (red solid line) and $\mu=\mu_{\mathrm{CEP}}$ (green solid line) at vanishing angular velocity. Gray dashed lines show the unstable and metastable branches of the solutions.}
\label{speIo0}
\end{figure}

For comparison, Fig.\,\ref{speIo0} show the same quantities at vanishing angular velocity for several values of the chemical potential, i.e., to $\mu=0$, $0.2~\mathrm{GeV}$, and $\mu=\mu_{\mathrm{CEP}}$ represented by the solid blue, red, and green curves, respectively. The insets focus on the neighborhood of the transition temperature in order to make the evolution of the curves with increasing chemical potential more transparent. As the chemical potential approaches its critical value, the coexistence region gradually shrinks. At $\mu=\mu_{\mathrm{CEP}}$, the three branches merge into a single solution and the multivalued structure disappears, signaling the onset of a second-order phase transition. 
The overall effect of increasing the chemical potential is found to be qualitatively similar to that produced by increasing the angular velocity. In particular, larger values of $\mu$ also lead to an enhancement of the scaled entropy density, pressure, energy density, and trace anomaly. In this respect, our results do not support a sharp distinction between the effects of rotation and chemical potential at the level of these thermodynamic observables. The main difference appears in the high-temperature region, where the asymptotic values remain essentially unchanged as $\mu$ varies, whereas they exhibit a noticeable dependence on the angular velocity.

Although the modifications induced by $\omega$ and $\mu$ share several common features, the origin of these effects is not immediately apparent from the thermodynamic quantities themselves. A more detailed understanding emerges from the thermodynamic response functions and the resulting phase diagrams, to which we now turn.

\subsection{Thermodynamic response functions}

To characterize the response of the rotating plasma to thermal and density fluctuations, it is useful to consider thermodynamic quantities involving derivatives of the grand potential. Among these, the specific heat at fixed chemical potential, the baryon number susceptibility, and the speed of sound are particularly sensitive to changes in the phase structure and therefore provide useful probes of the confinement--deconfinement transition.

\begin{figure}[h]
\begin{center}
\includegraphics[width=6.8cm]{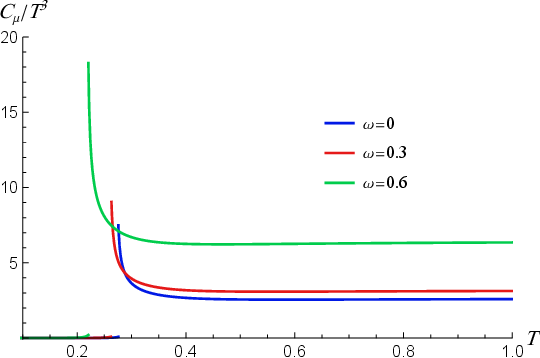}\hspace{.3cm}
\includegraphics[width=6.8cm]{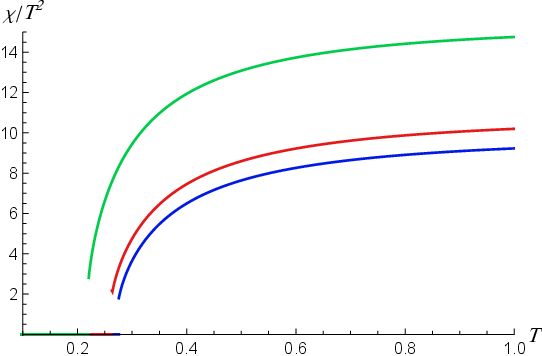}\\\vspace{.2cm}\includegraphics[width=6.8cm]{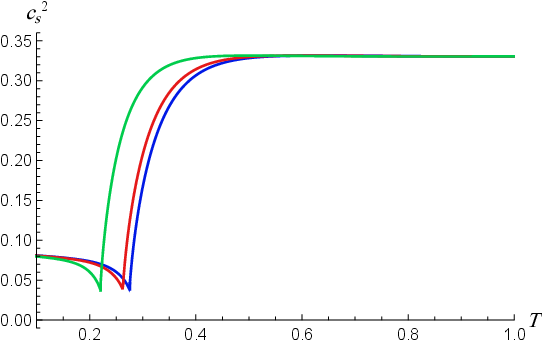}
\end{center}
\caption{\footnotesize 
Thermodynamic response functions $C_\mu/T^3$, $\chi/T^2$, and $c_s^2$ for various values of angular velocity $\omega=0$ (blue line), $\omega=0.3\,\mathrm{GeV}$ (red line) and $\omega=0.6\,\mathrm{GeV}$ (green line) at vanishing chemical potential.}
\label{responsemu0}
\end{figure} 

The specific heat at fixed chemical potential is defined by
\begin{align}
C_\mu=T\left(\frac{\partial s}{\partial T}\right)_{\mu,\omega},
\label{Cmudef}
\end{align}
and measures the response of the entropy density to variations of the temperature. The baryon number susceptibility is given by
\begin{align}
\chi=\left(\frac{\partial \rho}{\partial \mu}\right)_{T,\omega},
\label{chidef}
\end{align}
and quantifies density fluctuations in the medium. The squared speed of sound is obtained from
\begin{align}
c_s^2=\left(\frac{\partial p}{\partial \epsilon}\right)_{\mu,\omega},
\label{csdef}
\end{align}
which can equivalently be evaluated from the thermodynamic relations discussed above. In holographic QCD models, these quantities are known to provide sensitive indicators of critical behavior and are widely used to locate first-order, second-order, and crossover transitions.

The behavior of the scaled response functions $C_\mu/T^3$, $\chi/T^2$, and $c_s^2$ for several values of the angular velocity, $\omega=0$, $0.3~\mathrm{GeV}$, and $0.6~\mathrm{GeV}$, is shown in Fig.\,\ref{responsemu0}, where the stable solutions are represented by the solid blue, red, and green curves, respectively. As the angular velocity increases, all three quantities are enhanced throughout the intermediate-temperature region. For $C_\mu/T^3$ and $\chi/T^2$, this enhancement persists into the high-temperature regime, where the asymptotic values increase with $\omega$. In contrast, the speed of sound approaches constant limiting values at both low and high temperatures which remain essentially insensitive to the angular velocity. The minimum value of $c_s^2$, occurring in the vicinity of the transition region, is also found to be nearly independent of $\omega$ and remains positive and close to zero for all angular velocities considered here.

\begin{figure}[h]
\begin{center}
\includegraphics[width=6.8cm]{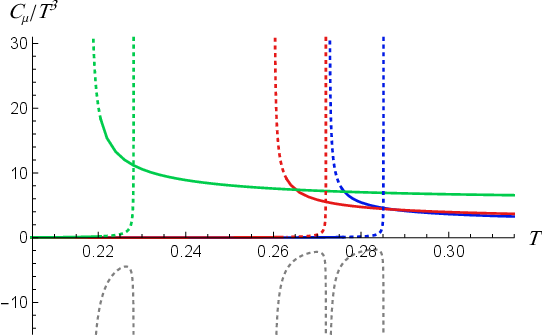}\hspace{.3cm}
\includegraphics[width=6.8cm]{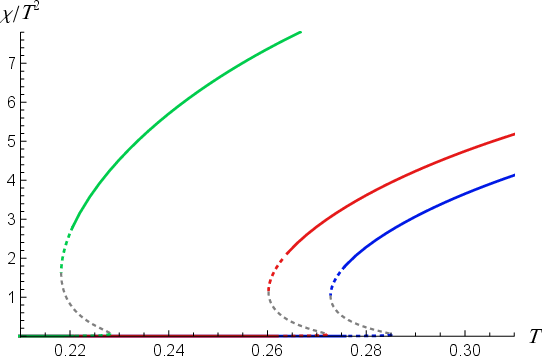}\\\vspace{.2cm}\includegraphics[width=6.8cm]{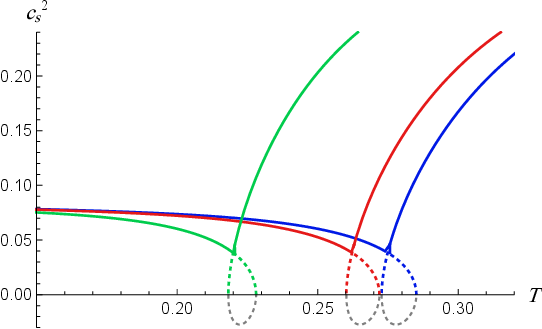}
\end{center}
\caption{\footnotesize 
The stable and metastable solutions of the rescaled thermodynamic response functions for various values of angular velocity $\omega=0$ (solid and dashed blue lines), $\omega=0.3\,\mathrm{GeV}$ (solid and dashed red lines) and $\omega=0.6\,\mathrm{GeV}$ (solid and dashed green lines) at vanishing chemical potential, and the unstable solutions shown by dashed gray curves in all cases.}
\label{responsemu0unstable}
\end{figure} 

\begin{figure}[h]
\begin{center}
\includegraphics[width=6.8cm]{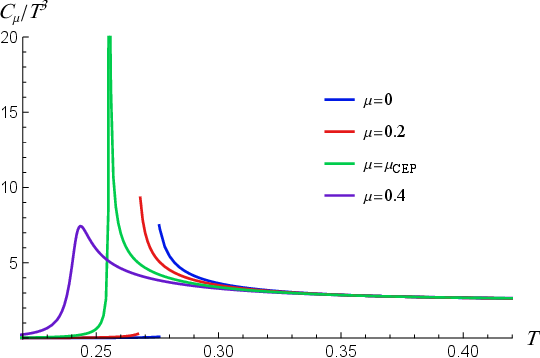}\hspace{.3cm}
\includegraphics[width=6.8cm]{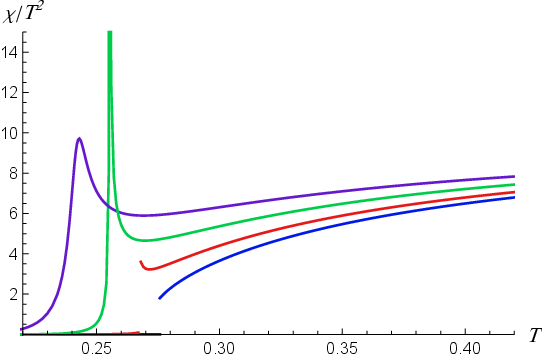}\\\vspace{.2cm}\includegraphics[width=6.8cm]{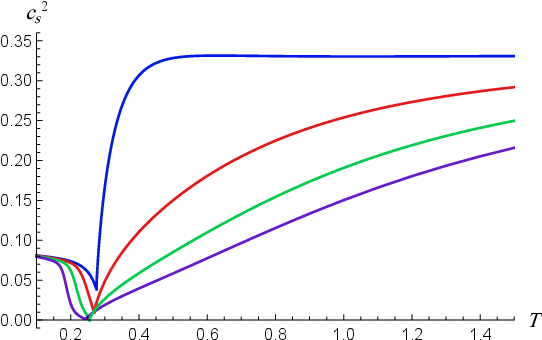}
\end{center}
\caption{\footnotesize 
Thermodynamic response functions $C_\mu/T^3$, $\chi/T^2$, and $c_s^2$ for various values of chemical potential $\mu=0$ (blue line), $\mu=0.2\,\mathrm{GeV}$ (red line), $\mu=\mu_{\mathrm{CEP}}$ (green line) and $\mu=0.4\,\mathrm{GeV}$ (purple line) at vanishing angular velocity.}
\label{responseo0}
\end{figure}

To make the branch structure more transparent, Fig.\,\ref{responsemu0unstable} displays enlarged views of the transition region including the stable, metastable, and unstable solutions. The stable branches are shown by the solid blue, red, and green curves, while the corresponding metastable extensions are represented by dashed curves of the same colors. The unstable branch is shown by the dashed gray curves. 
The response functions exhibit distinct signatures of the first-order transition. The specific heat $C_\mu$ develops singularities at the spinodal boundaries separating the stable, metastable, and unstable branches. The unstable branch is characterized by a negative specific heat and therefore does not correspond to a thermodynamically realized state. In contrast, the baryon number susceptibility $\chi$ remains finite but becomes multivalued within the coexistence region, reflecting the existence of multiple competing black-hole solutions at the same temperature.
The behavior of the speed of sound provides a particularly transparent characterization of the stability properties of the different branches. The metastable solutions remain associated with positive values of $c_s^2$, which decrease toward zero near the spinodal boundaries. The unstable branch connecting them is characterized by $c_s^2<0$, signaling mechanical instability. As a consequence, the vanishing of $c_s^2$ marks the boundaries between the metastable and unstable regions.

The effect of the chemical potential is illustrated in Fig.\,\ref{responseo0}, where the same observables are shown at vanishing angular velocity for several values of $\mu$. The blue, red, and green curves correspond to $\mu=0$, $0.2~\mathrm{GeV}$, and $\mu=\mu_{\mathrm{CEP}}$, respectively, while the violet curve represents $\mu=0.4~\mathrm{GeV}$ which is beyond the critical end point. Similar to the angular velocity, increasing $\mu$ enhances all three response functions in the intermediate-temperature region. Unlike the rotational case, however, the asymptotic high-temperature values remain essentially unchanged as the chemical potential varies.

\begin{figure}[h]
\begin{center}
\includegraphics[width=6.8cm]{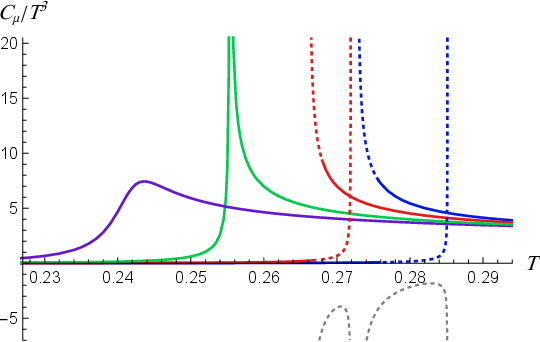}\hspace{.3cm}
\includegraphics[width=6.8cm]{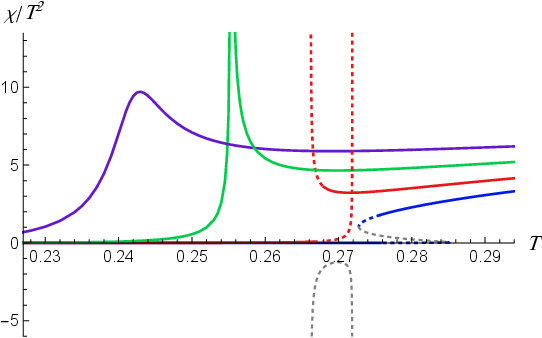}\\\vspace{.2cm}\includegraphics[width=6.8cm]{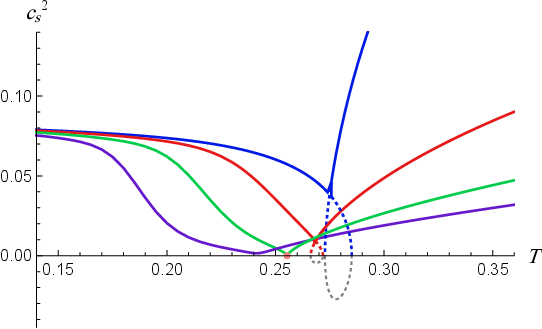}
\end{center}
\caption{\footnotesize 
All the branches of the rescaled thermodynamic response functions for various values of the chemical potential at vanishing angular velocity. For $\mu=0$ and $\mu=0.2\,\mathrm{GeV}$ which are below the critical end point, the stable and metastable branches are respectively depicted by solid and dashed blue lines, and solid and dashed red lines. The unstable solutions are also shown by the dashed gray curves. The stable solutions for $\mu=\mu_{\mathrm{CEP}}$ and $\mu=0.4\,\mathrm{GeV}>\mu_{\mathrm{CEP}}$ are respectively shown by solid green and solid purple lines.}
\label{responseo0unstable}
\end{figure}

A closer view of the transition region is presented in Fig.\,\ref{responseo0unstable}, where the branch structure is shown using the same color convention as in Fig.\,\ref{responsemu0unstable}. As the chemical potential increases, the coexistence region gradually shrinks and the stable, metastable, and unstable branches move toward each other. 
The response functions exhibit distinct signatures of this evolution. The specific heat develops spinodal singularities throughout the first-order region, while the baryon number susceptibility shows a more subtle behavior. At vanishing chemical potential, when there is no finite background density, $\chi$ remains finite and exhibits a multivalued structure within the coexistence region. For nonzero chemical potentials, however, the susceptibility develops divergences at the spinodal boundaries and becomes negative along the unstable branch, reflecting the loss of thermodynamic stability associated with density fluctuations.
For both of these response functions, at the critical end point, the two singularities merge into a single critical divergence, signaling the onset of a second-order phase transition. 
The behavior of the speed of sound provides an additional characterization of the branch structure. As the chemical potential approaches the critical value, the metastable and unstable branches progressively shrink and the minimum of $c_s^2$ decreases. At the critical end point, all branches merge into a single solution and the minimum of the speed of sound reaches zero, signaling the onset of critical behavior associated with the second-order phase transition. For larger chemical potentials, the branch structure disappears entirely and all response functions evolve smoothly across the transition region, indicating a crossover. The evolution of these observables therefore provides a clear thermodynamic picture of how the first-order transition terminates at the critical end point and subsequently turns into a crossover. 

The behavior of the response functions provides a set of precise thermodynamic criteria for locating the phase boundaries of the rotating plasma. In the next subsection, we employ these observables together with the thermodynamic quantities discussed above to construct the phase diagrams in the $(T,\mu)$ and $(T,\omega)$ planes. This allows us to investigate the influence of rotation on the confinement--deconfinement transition, the location of the critical end point, and the crossover region beyond criticality.

\subsection{Phase diagrams and scaling properties}

\begin{figure}[h]
\begin{center}
\includegraphics[width=6.8cm]{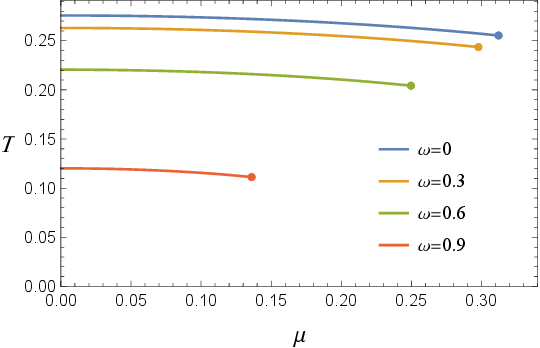}\hspace{.3cm}
\includegraphics[width=6.8cm]{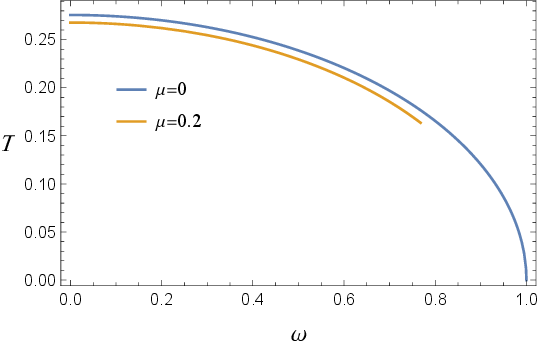}
\end{center}
\caption{\footnotesize 
Left graph: The phase diagram in the $(T,\mu)$ plane extracted from the swallow tail structure of the grand potential for various values of the angular velocity. Right graph: The phase diagram in the $(T,\omega)$ plane extracted from the swallow tail structure of the grand potential for $\mu=0$ and $\mu=0.2\,\mathrm{GeV}$. Solid lines show the first-order transition and dots indicate the CEP.}
\label{phasediagramsOmega}
\end{figure} 

Having analyzed the thermodynamic observables and response functions, we now determine the phase structure of the rotating plasma. The phase boundaries are obtained from the behavior of the grand potential and the response functions discussed in the previous subsections. Particular attention will be paid to the study of the influence of the angular velocity on the phase diagram.

The phase diagrams in the $(T,\mu)$ plane extracted from the grand potential are shown in the left graph of Fig.\,\ref{phasediagramsOmega} for several values of the angular velocity. The solid curves represent the first-order confinement--deconfinement transition, while the dots indicate the corresponding critical end points. As the angular velocity increases, the entire transition line is shifted toward lower temperatures and chemical potentials. Consequently, both coordinates of the critical end point decrease monotonically with increasing $\omega$.

\begin{figure}[h]
\begin{center}
\includegraphics[width=6.8cm]{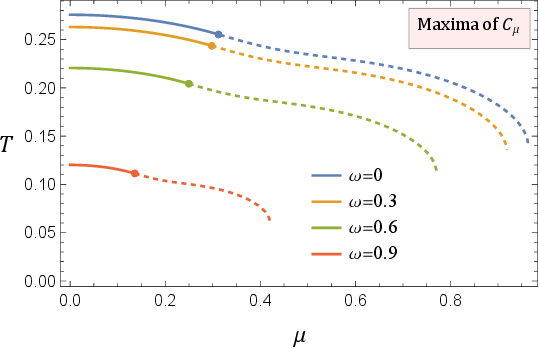}\hspace{.3cm}
\includegraphics[width=6.8cm]{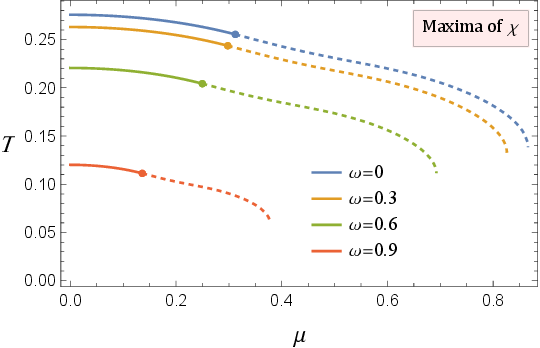}\\\vspace{.2cm}\includegraphics[width=6.8cm]{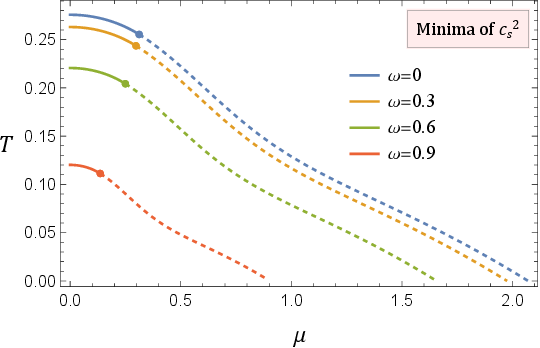}
\end{center}
\caption{\footnotesize 
The phase diagram in the $(T,\mu)$ plane including the separation lines in the supercritical region, obtained using the extrema of the response functions $C_\mu$, $\chi$, and $c_s^2$.}
\label{Tmuresponse}
\end{figure} 

The same information can be represented in the $(T,\omega)$ plane at fixed chemical potential, as shown in the right graph of  Fig.\,\ref{phasediagramsOmega}. For $\mu=0$, the first-order transition line extends from $\omega=0$ to $\omega=1\,\mathrm{GeV}$, where the transition temperature vanishes. For nonzero chemical potential, however, the transition line terminates at smaller values of the angular velocity. As the chemical potential approaches the critical value, the first-order region continuously shrinks and disappears at $\mu=\mu_{\rm CEP}$, beyond which no first-order transition survives.

\begin{figure}[h]
\begin{center}
\includegraphics[width=6.8cm]{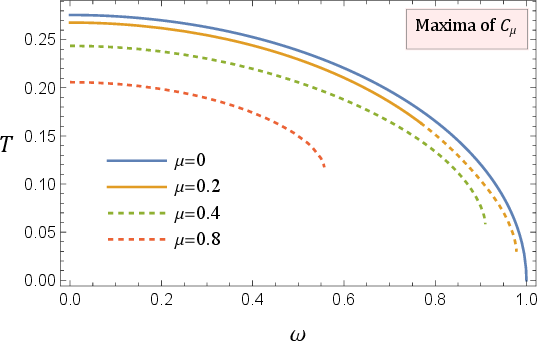}\hspace{.3cm}
\includegraphics[width=6.8cm]{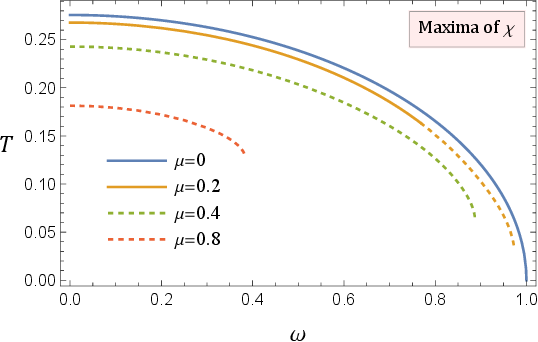}\\\vspace{.2cm}\includegraphics[width=6.8cm]{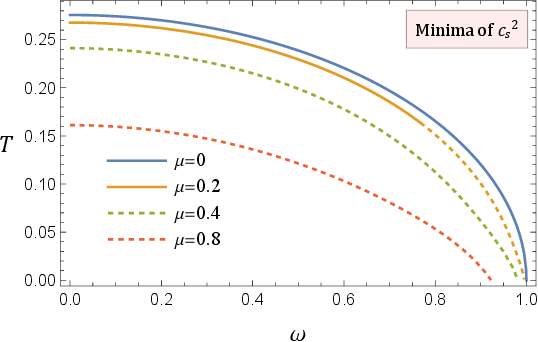}
\end{center}
\caption{\footnotesize 
The phase diagram in the $(T,\omega)$ plane including the separation lines in the supercritical region, obtaind using the extrema of the response functions $C_\mu$, $\chi$, and $c_s^2$.}
\label{Tomegaresponse}
\end{figure}

To investigate the supercritical region, we extend the phase diagrams beyond the critical end point using the extrema of the response functions introduced above. The resulting diagrams obtained from $C_\mu$, $\chi$, and $c_s^2$ are shown in Figs.\,\ref{Tmuresponse} and \ref{Tomegaresponse}. 
In each case, the solid curves denote the first-order transition line, while the dashed curves represent the crossover lines determined from the extrema of the corresponding response function. As already observed in the static model \cite{superus}, different thermodynamic probes lead to different crossover trajectories in the supercritical region. This reflects the absence of a genuine phase transition beyond the critical end point and the consequent ambiguity in defining a unique crossover line. Nevertheless, all crossover curves originate from the critical end point and provide a consistent characterization of the gradual change between the confined and deconfined regimes.

A remarkable feature emerging from all thermodynamic observables and phase diagrams is the simple manner in which the angular velocity enters the thermodynamic sector of the theory. Since the rotating background is generated from the static solution through the boost relations
\begin{align}
T_0=\gamma T,
\qquad
\mu_0=\gamma\mu,
\end{align}
it is natural to expect a corresponding scaling structure in the thermodynamic observables. Indeed, the numerical results indicate that the thermodynamic sector is accurately described by the scaling relation
\begin{align}
\Omega(T,\mu,\omega)=\Omega_0(\gamma T,\gamma\mu),
\label{omegascaling}
\end{align}
where $\Omega_0$ denotes the grand potential of the static background. Since all thermodynamic observables and response functions can be obtained from derivatives of the grand potential, Eq.\,\eqref{omegascaling}
immediately determines their rotational dependence. 
The phase boundaries therefore inherit the same scaling structure. If $T_c(\mu,0)$ denotes a transition or crossover line in the static background, the corresponding line in the rotating geometry satisfies
\begin{align}
T_c(\mu,\omega)=\frac{T_c(\gamma\mu,0)}{\gamma}.
\end{align}
This relation is found to hold for all thermodynamic probes considered in this work, including the first-order transition line and the
crossover trajectories in the supercritical region. In particular, the coordinates of the critical end point obey
\begin{align}
T_{\rm CEP}(\omega)=\frac{T_{\rm CEP}(0)}{\gamma},
\qquad
\mu_{\rm CEP}(\omega)=\frac{\mu_{\rm CEP}(0)}{\gamma},
\end{align}
in excellent agreement with the numerical results.

 Together, these results indicate that the rotating construction employed here does not generate an independent thermodynamic deformation of the EMD background. Instead, the entire thermodynamic sector of the rotating plasma can be mapped onto that of the static system through the boost-induced rescaling of the temperature and chemical potential. The influence of rotation on thermodynamic observables is therefore largely kinematical and can be understood as a consequence of the Lorentz transformation used to construct the rotating geometry.

This observation raises an important question. If the thermodynamic sector is completely determined by the corresponding static solution through a simple rescaling, can the rotating geometry nevertheless produce genuinely new physical effects? To address this issue, it is necessary to consider observables probing the bulk geometry more directly. In the next section, we therefore turn to the Schwinger effect and investigate whether the response of the system to an external electric field exhibits signatures of rotation beyond the thermodynamic scaling discussed above.

\section{Schwinger effect and string probes of the rotating plasma}

In holography, quantities associated with fundamental strings provide access to aspects of the strongly coupled plasma that are not encoded solely in equilibrium thermodynamics. In particular, the Schwinger effect and the quark--antiquark potential are sensitive to the structure of the string worldsheet and therefore furnish a useful probe of the rotating geometry.

In this section, we investigate the influence of the angular velocity on the Schwinger effect using the potential analysis. We first construct the string configuration in the rotating background and determine the effective string tension governing the dynamics of the quark--antiquark pair. This allows us to identify the infrared wall associated with confinement and to examine how the resulting confinement criterion is modified by the rotation. We then analyze the behavior of the string configurations in both confined and deconfined phases and study the corresponding critical electric fields and total potential.


\subsection{String configuration and effective string tension}

To investigate the Schwinger effect, we employ the potential analysis \cite{potential}. The quantity of interest is the total potential energy of a static quark--antiquark pair immersed in the rotating plasma in the presence of an external electric field.

Since the rotating geometry is generated through a compactification of the $x_3$ direction followed by a boost in the resulting angular coordinate $\theta$, the spatial directions are no longer equivalent. In an anisotropic medium, it is in general interesting to distinguish between quark--antiquark pairs oriented parallel and perpendicular to the rotation axis. The present construction, however, naturally restricts the class of configurations that can be investigated.

Indeed, the rotating plasma is described on a slice of fixed distance $l$ from the rotation axis. As a consequence, the radial direction within the rotation plane is absent from the effective boundary geometry and cannot be used to define the quark--antiquark separation. The remaining direction in the rotation plane is the compact angular coordinate $\theta$, which is precisely the direction employed in constructing the rotating background. Since $\theta$ is compact and already mixed with time through the boost transformation, a Wilson line extended along this direction acquires a qualitatively different interpretation: it probes a co-rotating angular separation rather than a pure spatial interval.
Consequently the boost construction employed here cannot capture the full parallel-versus-perpendicular anisotropy that is accessible in cylindrical rotating backgrounds; we therefore restrict the analysis to the parallel orientation.
In the present work, we consider a quark--antiquark pair separated along one of the directions perpendicular to the compactified circle, chosen without loss of generality to be $x_1$, while keeping its position fixed in the $\theta$ and $x_2$ directions. From the boundary-theory perspective, this corresponds to a pair located at a fixed distance $l$ from the rotation axis and separated along a direction parallel to the rotation axis. The pair therefore co-rotates with the plasma while maintaining a constant spatial separation $L$.

Within the potential analysis, the total potential energy of the quark--antiquark pair in the presence of an external electric field is obtained from the dynamics of an open string whose endpoints are attached to a probe D3-brane located at a finite radial position $z_0$. The position of the probe brane determines the constituent quark mass and prevents the appearance of infinitely heavy quarks.
The relevant string configuration is therefore described by the static gauge
\begin{align}
t=\tau,
\qquad
x_1=\sigma,
\qquad
z=z(\sigma),
\label{staticgauge}
\end{align}
with all remaining coordinates held fixed. In the following, for notational simplicity, we denote the spatial coordinate along the separation of the quark pair by $x$.

The dynamics of the string is governed by the Nambu--Goto action
\begin{align}
S_{\rm NG}=T_F \int d\tau d\sigma
\sqrt{-\det g_{\alpha\beta}},
\label{NGaction}
\end{align}
where $T_F=\frac{1}{2\pi\alpha'}$ denotes the fundamental string tension and $g^s_{\alpha\beta}=g^s_{\mu\nu}\frac{\partial X^\mu}{\partial\xi^\alpha}\frac{\partial X^\nu}{\partial\xi^\beta}$ is the string-frame metric induced on the worldsheet. Here, $\xi^\alpha$ denote the  worldsheet coordinates, while $X^\mu$ and $g^s_{\mu\nu}$  are the bulk coordinates and the string-frame metric, respectively.

Substituting the rotating metric \eqref{rotatingmetric} in the string frame into Eq.\,\eqref{NGaction}, one finds
\begin{align}
S_{\rm NG}=T_F {\cal T} \int_{-L/2}^{L/2} dx \sqrt{\frac{k(z)}{G(z)}} \sqrt{z'(x)^2+G(z)},
\label{NGaction2}
\end{align}
where ${\cal T}=\int d\tau$  is taken over a sufficiently long time interval and $k(z)=\left[N(z)-R(z) P(z)^2\right]H_s(z)^2$. Here, $H_s(z)=\frac{e^{2A_s(z)}}{z^2}$ and $A_s(z)=A_e(z)+\sqrt{\frac{1}{6}}\phi(z)$.

To determine the potential energy of the quark-antiquark pair separated by a distance $L$, one must solve $\frac{S_{\rm NG}}{{\cal T} }$ for a U-shaped open string with profile $z(x)$. The string endpoints are attached to the probe D3-brane at $x=-\frac{L}{2}$ and $x=\frac{L}{2}$, while its turning point in the radial direction is denoted by $z_c$. The profile of the string should satisfy the conditions $z(0)=z_c$ and $z'(0)=0$. 

Since the Lagrangian does not depend explicitly on $x$, there exists a conserved quantity along the string worldsheet,
\begin{align}
{\cal H}=\frac{\sqrt{k(z)G(z)}}{\sqrt{z'(x)^2+G(z)}}=\sqrt{k(z_c)}. 
\end{align}
Integrating this relation, the separation length between the quark and antiquark is obtained as
\begin{align}
L=2\int_{z_0}^{z_c}dz\,\frac{\sqrt{k(z_c)}}{\sqrt{G(z)}\sqrt{k(z)-k(z_c)}}.
\label{length}
\end{align}
Similarly, the sum of the potential and static energies associated with the connected string configuration is given by
\begin{align}
V_{\rm PE+SE}
=2T_F
\int_{z_0}^{z_c}
dz\,\frac{k(z)}{\sqrt{G(z)}\sqrt{k(z)-k(z_c)}}.
\label{potential}
\end{align}
The total potential is then obtained by including the interaction with a constant external electric field $E$,
\begin{align}
V_{\rm tot}=V_{\rm PE+SE}-EL,
\end{align}
 in which the separation length is given by Eq.\,\eqref{length}. The second term tends to pull the quark and antiquark apart and consequently lowers the potential barrier responsible for vacuum stability.

Before proceeding to the numerical analysis, it is useful to discuss the radial region of the bulk that is accessible to the string configuration. This issue plays a central role in the potential analysis since both the separation length \eqref{length} and the potential \eqref{potential} are determined by integrations over the radial coordinate. The existence of special points that restrict the motion of the string therefore directly affects the behavior of the quark--antiquark potential and the Schwinger effect.

The confinement properties of the background can be conveniently characterized through the effective string tension \cite{superus}
\begin{align}
\sigma_{\rm eff}(z)
=\sqrt{-g^s_{tt}(z)\,g^s_{xx}(z)}=\sqrt{k(z)}.
\label{effectiveST}
\end{align}
In confining geometries, $\sigma_{\rm eff}(z)$ develops a nonvanishing minimum at a finite radial position, $z=z_{\rm IR}$, which corresponds to a dynamical infrared wall. As the turning point of the string approaches this position, the denominator of Eq.\,\eqref{length} vanishes and the separation length diverges.
Consequently, the string cannot penetrate beyond $z_{\rm IR}$ and the accessible radial region is restricted to
\begin{align}
z_0\leq z\leq z_{\rm IR}.
\end{align}
From the boundary-theory perspective, the existence of such an infrared wall signals confinement since an arbitrarily large separation between the quark and antiquark requires an infinite amount of energy.

The value of the effective string tension at the minimum determines the confining string tension,
\begin{align}
\sigma_s=\sigma_{\rm eff}(z_{\rm IR}).
\end{align}
It also fixes the lowest critical electric field required to destabilize the confining vacuum,
\begin{align}
E_s=T_F\sigma_s,
\end{align}
which constitutes one of the characteristic scales governing the Schwinger effect in the confined phase.

In the deconfined phase, on the other hand, the infrared wall disappears and the separation length remains finite for all connected string configurations. In the static background, the deeper branch of the U-shaped string extends continuously toward the black-hole horizon, and the turning point can approach $z_h$ arbitrarily closely. The rotating geometry, however, exhibits a novel feature.

To understand this point, let us consider the induced metric on the string worldsheet. For the static embedding \eqref{staticgauge}, by using the rotating metric \eqref{rotatingmetric}, the temporal component of the induced metric is simply
\begin{align}
g^{\rm ws}_{\tau\tau}=g_{tt}\propto
-\gamma^2\left[G(z)-\omega^2l^2\right].
\end{align}
Consequently, there exists a special radial position $z_*$ determined by
\begin{align}
G(z_*)=\omega^2l^2,
\label{zstarcondition}
\end{align}
at which the Killing vector $\partial_t$ becomes null on the worldsheet. This point therefore corresponds to a worldsheet horizon which is equivalent to the critical surface fixed by the reality condition of the Nambu-Goto action.
The position $z_*$ satisfies $0\leq z_*\leq z_h$ and reduces to the black-hole horizon in the static limit, $z_*(\omega=0)=z_h$.
As a consequence, the connected string solutions are restricted to
\begin{align}
z_0\leq z\leq z_*,
\end{align}
rather than the interval extending all the way to the event horizon. In other words, the worldsheet horizon replaces the black-hole horizon as the deepest radial position accessible to the connected string configuration. 

The appearance of such a worldsheet horizon is familiar from studies of moving probes and trailing strings in strongly coupled plasmas, where it plays a central role in determining the dynamics of the string \cite{zstar1,zstar2,zstar3,zstar4,zstar5,zstar6}. As we shall show below, an analogous mechanism operates in the present rotating setup. Increasing the angular velocity pushes the worldsheet horizon toward the boundary, thereby reducing the portion of the bulk geometry accessible to the string and modifying the behavior of the quark--antiquark potential, which constitutes one of the distinctive dynamical consequences of the boost-generated rotation.

\subsection{Infrared wall and phase structure}

The existence of a dynamical infrared wall provides a simple criterion for distinguishing between the confined and deconfined phases from the perspective of the string degrees of freedom. Whenever the effective string tension possesses a minimum at $z=z_{\rm IR}$, the separation length diverges as $z_c\rightarrow z_{\rm IR}$, signaling confinement. The disappearance of this minimum indicates that the system has entered the deconfined phase. The behavior of $z_{\rm IR}$ therefore encodes valuable information about the phase structure experienced by a quark--antiquark pair.

\begin{figure}[h]
\begin{center}
\includegraphics[width=6.8cm]{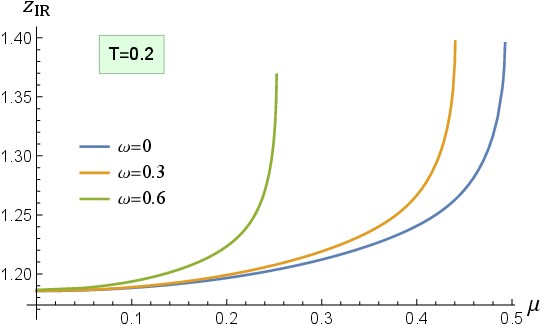}\hspace{.3cm}
\includegraphics[width=6.8cm]{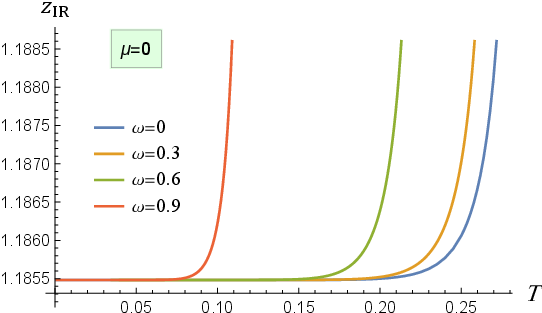}\\\vspace{.2cm}\includegraphics[width=6.8cm]{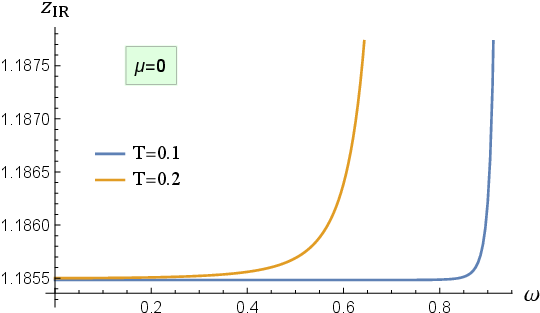}
\end{center}
\caption{\footnotesize 
The dependence of $z_{\rm IR}$ on different thermodynamic parameters.}
\label{zIRplots}
\end{figure} 

The dependence of the infrared wall position on the thermodynamic parameters is shown in Fig.\,\ref{zIRplots}. The top left panel displays $z_{\rm IR}$ as a function of the chemical potential at fixed temperature $T=0.2~{\rm GeV}$ for several values of the angular velocity. For all values of $\omega$, the infrared wall moves deeper into the bulk as the chemical potential increases. The effect of rotation follows the same trend, although it remains quantitatively small. At low chemical potentials the curves corresponding to different angular velocities are close to each other, while their separation becomes more noticeable at larger values of $\mu$, indicating that the influence of rotation is enhanced in denser plasmas.

The top right panel shows the temperature dependence of the infrared wall at vanishing chemical potential. As expected, increasing the temperature gradually increases $z_{\rm IR}$ or equivalently pushes the infrared wall toward the black-hole horizon. 
However, the infrared wall always remains outside the black-hole horizon, i.e., $z_{\rm IR}<z_h$. $z_{\rm IR}$ does not merge smoothly with the $z_h$ when approaching the confinement--deconfinement transition. Regardless of the value of the angular velocity, $z_{\rm IR}$ approaches the transition with a finite value noticeably smaller than $z_h$ and disappears abruptly once the transition is reached. The effect of the angular velocity is again rather mild, with larger values of $\omega$ shifting the wall slightly deeper into the bulk. An interesting feature of the figure is that all curves approach the same limiting value as $T\rightarrow0$, independently of the angular velocity. This behavior differs from that observed when varying the chemical potential, where distinct values of $\mu$ generally lead to different zero-temperature limits, consistent with the presence of charged extremal black-hole solutions \cite{superus}. 

The dependence of $z_{\rm IR}$ on the angular velocity is illustrated in the third panel of Fig.\,\ref{zIRplots} for two representative temperatures at $\mu=0$. The infrared wall moves monotonically toward larger values as the angular velocity increases, indicating that rotation tends to weaken the confining behavior seen by the string probe. Nevertheless, the magnitude of the effect remains small throughout the parameter range considered here.

\begin{figure}[h]
\begin{center}
\includegraphics[width=6.8cm]{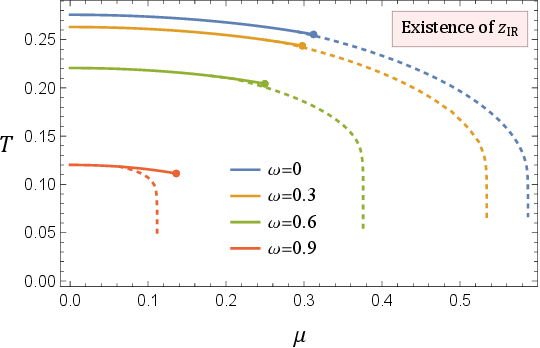}\hspace{.3cm}
\includegraphics[width=6.8cm]{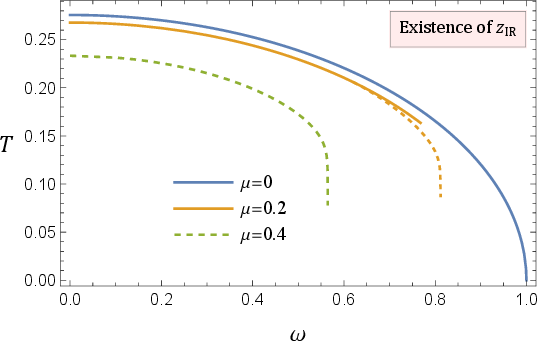}
\end{center}
\caption{\footnotesize 
The phase diagram in the $(T,\mu)$ (left) and $(T,\omega)$ (right) planes. The solid curves denote the first-order transition lines obtained from the grand potential, while the dashed curves in the supercritical region are determined from the string confinement criterion, namely the appearance/disappearance of the dynamical infrared wall $z_{\rm IR}$.}
\label{phasediagramszw}
\end{figure} 

Since confinement from the string perspective is directly tied to the existence of the infrared wall, the confinement--deconfinement transition can be determined by identifying the points in parameter space at which $z_{\rm IR}$ disappears. The resulting phase diagrams in the $(T,\mu)$ and $(T,\omega)$ planes are shown in Fig.\,\ref{phasediagramszw}. The solid curves represent the first-order transition lines previously obtained from the grand potential, while the dashed curves correspond to the transition lines extracted from the infrared-wall criterion.

As in the static EMD model studied in \cite{superus}, the two definitions of confinement do not coincide over the entire first-order region. Starting from sufficiently large values of the chemical potential or angular velocity, the transition line determined by the infrared wall separates from the thermodynamic first-order transition line and extends into the supercritical region. Consequently, a finite domain appears in which the system is thermodynamically confined according to the grand potential and other thermodynamic probes, while the string description already exhibits deconfined behavior. This mismatch reflects the fact that the thermodynamic and string probes are sensitive to different aspects of the underlying dynamics.

\begin{figure}[h]
\begin{center}
\includegraphics[width=6.8cm]{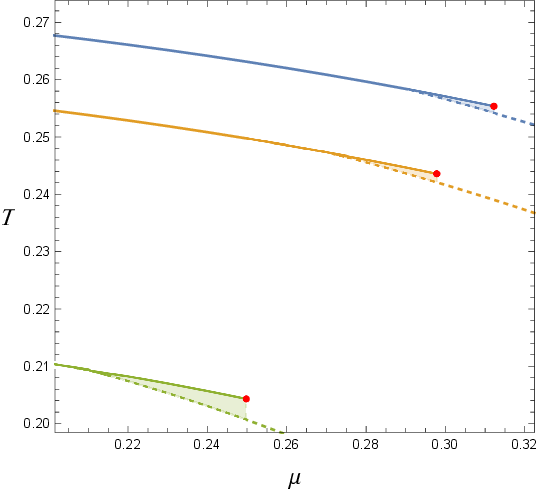}
\end{center}
\caption{\footnotesize 
Enlarged views of the mismatch regions of confinement in the phase diagram in the $(T,\mu)$ plane in Fig\,\ref{phasediagramszw}.}
\label{triangularregion}
\end{figure}

The enlarged views shown in Fig.\,\ref{triangularregion} make this feature more transparent. The shaded regions enclosed between the two transition lines correspond precisely to the domain where the two criteria lead to different conclusions regarding confinement. The most notable effect of rotation is that this region grows with increasing angular velocity. In the $(T,\mu)$ plane, the separation between the two curves occurs at progressively smaller values of the chemical potential as $\omega$ increases, leading to a larger mismatch region. Rotation therefore enlarges the parameter space in which the plasma remains thermodynamically confined while the string degrees of freedom behave as if they were already in a deconfined phase.

The phase diagrams obtained from the infrared-wall criterion also reveal an important difference with respect to the thermodynamic observables discussed in Sec.\,3. There, all phase boundaries were found to obey the simple rescaling relations associated with the boost-generated rotation. No such scaling is observed for the confinement lines extracted from $z_{\rm IR}$. The corresponding curves for different values of $\omega$ cannot be mapped into one another through a simple Lorentz rescaling of the temperature and chemical potential. The string sector therefore exhibits a genuinely nontrivial dependence on the angular velocity that is absent in the thermodynamic observables. Nevertheless, the qualitative influence of rotation remains the same: increasing $\omega$ shifts the transition toward lower temperatures and chemical potentials and moves the corresponding critical points to smaller values of both $T$ and $\mu$.

This distinction between the thermodynamic and string probes will become even more apparent in the following subsections, where we investigate the properties of the quark--antiquark potential and the Schwinger effect in both phases.

\subsection{Schwinger effect in the rotating plasma}

Having established the confinement--deconfinement structure from the viewpoint of the string, we now investigate how the rotating plasma responds to a constant external electric field. Within the potential analysis approach, the electric field competes with the attractive interaction between the quark and antiquark and modifies the potential barrier governing the Schwinger pair-production process. Depending on the magnitude of the applied field and on whether the system is in the confined or deconfined phase, the vacuum may remain stable, decay through quantum tunneling, or become catastrophically unstable.

Two characteristic electric fields naturally arise in this framework. The first one is the confining critical field $E_s$, which exists only when the background is confined from the string viewpoint. It is determined by the confining string tension,
\begin{align}
E_s=T_F\sigma_{\rm eff}(z_{\rm IR}),
\end{align}
and therefore vanishes simultaneously with the disappearance of the infrared wall. Physically, $E_s$ is independent of the mass of the produced quarks and represents the minimum electric field required for pair production in a confining background. For $E<E_s$, the confining string cannot be broken and no pair production occurs, even for massless quarks.
The second critical field is obtained from the DBI action of the probe D3-brane located at $z_0$,
\begin{align}
E_c=T_F\sigma_{\rm eff}(z_0),
\end{align}
and determines the onset of catastrophic vacuum instability for quarks with the mass specified by the position of the probe brane. Unlike $E_s$, this quantity is well defined in both confined and deconfined phases since it depends only on the local geometry at the position of the probe brane, whose radial location holographically determines the constituent quark mass. For $E_s<E<E_c$, pair production still occurs through quantum tunneling across a finite potential barrier. 

\begin{figure}[h]
\begin{center}
\includegraphics[width=6.8cm]{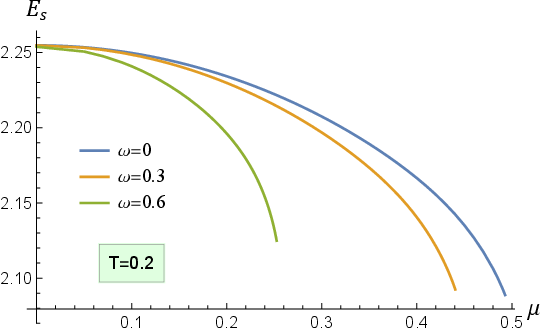}\hspace{.3cm}
\includegraphics[width=6.8cm]{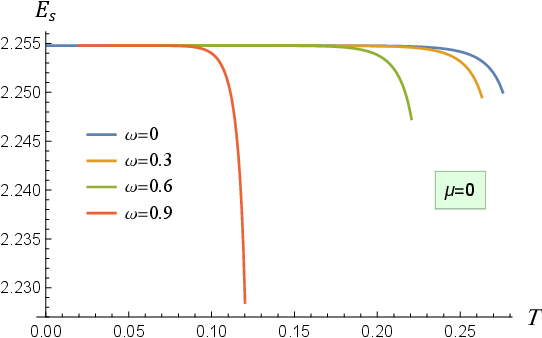}\\\vspace{.2cm}\includegraphics[width=6.8cm]{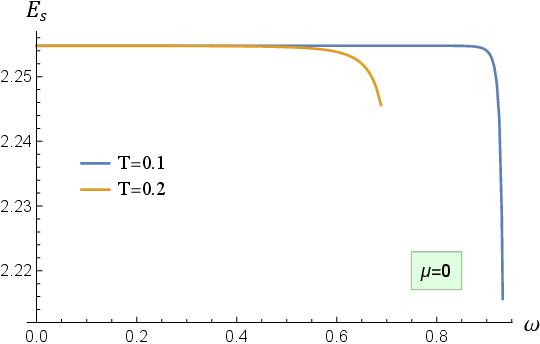}
\end{center}
\caption{\footnotesize 
The confining critical electric field $E_s$ as a function of the chemical potential (top left), temperature (top right), and angular velocity (bottom). The curves correspond to the values of $\omega$ (top panels) and $T$ (bottom panel) indicated in the legends.}
\label{Esplots}
\end{figure} 

The dependence of $E_s$ on the thermodynamic parameters is presented in Fig.\,\ref{Esplots}. The three panels display the variation of the confining critical field with the chemical potential, temperature and angular velocity, respectively. As expected from its definition, $E_s$ exists only when the infrared wall is present and therefore provides another direct probe of confinement from the quark perspective.

Increasing either the chemical potential or the temperature lowers the value of $E_s$, reflecting the progressive weakening of confinement. Rotation acts in exactly the same manner. For all values of the chemical potential and temperature considered here, increasing the angular velocity monotonically decreases the confining critical field. This behavior is consistent with the tendency of rotation to push the infrared wall deeper into the bulk, as shown in Fig.\,\ref{zIRplots}, thereby reducing the confining string tension.

\begin{figure}[h]
\begin{center}
\includegraphics[width=6.8cm]{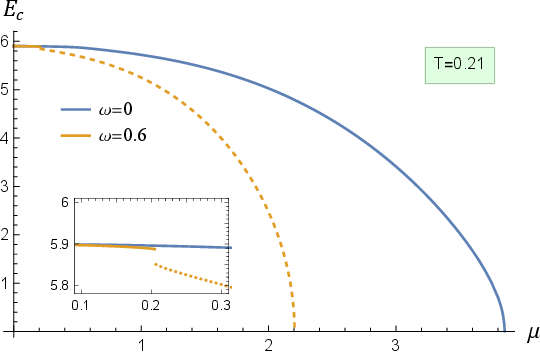}\hspace{.3cm}
\includegraphics[width=6.8cm]{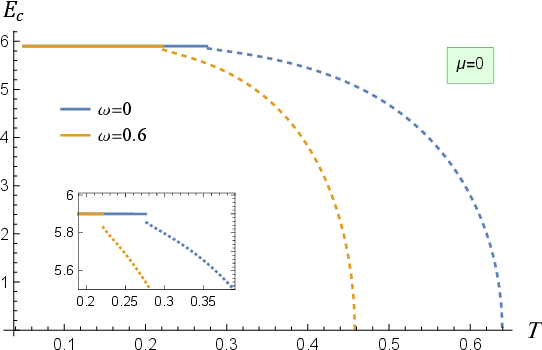}\\\vspace{.2cm}
\includegraphics[width=6.8cm]{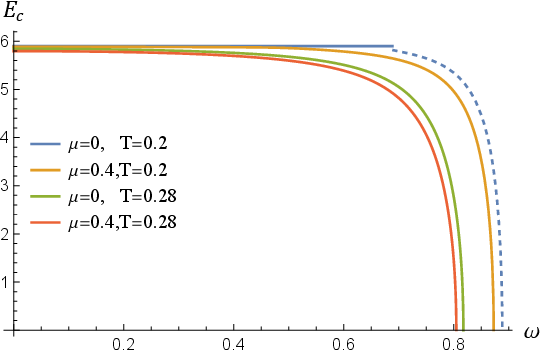}
\end{center}
\caption{\footnotesize 
The critical electric field $E_c$ as a function of the chemical potential (top left), temperature (top right), and angular velocity (bottom). Insets enlarge the discontinuities associated with first-order phase transitions. }
\label{Ecplots}
\end{figure} 

An interesting feature appears in the temperature dependence at vanishing chemical potential. Although different values of the angular velocity produce distinguishable curves at finite temperatures, they all approach the same limiting value as $T\rightarrow0$. Thus, unlike the chemical potential, whose influence persists at zero temperature through the charged extremal black-hole solution, the effect of rotation disappears in this limit. Moreover, the disappearance of $E_s$ precisely coincides with the disappearance of the infrared wall, confirming that the confining critical field is completely governed by the confinement criterion introduced in the previous subsection.

The corresponding behavior of the critical electric field $E_c$ is shown in Fig.\,\ref{Ecplots}. The qualitative dependence on the thermodynamic parameters is remarkably similar. Increasing the chemical potential, temperature or angular velocity always reduces the critical field, implying that a smaller external electric field is required to trigger catastrophic vacuum decay. From the experimental point of view, this tendency is particularly encouraging since one of the principal obstacles to observing the Schwinger effect in heavy-ion collisions is the extremely large electric field required for pair production. The reduction of $E_c$ induced by rotation therefore enhances the possibility of observing the Schwinger effect in a rotating quark--gluon plasma.

The behavior of $E_c$ across the phase transition, however, differs qualitatively from that of $E_s$. As illustrated by the enlarged views in Fig.\,\ref{Ecplots}, $E_c$ exhibits a finite discontinuity whenever the system crosses the first-order thermodynamic transition determined from the grand potential. In contrast, no analogous feature is observed when the trajectory crosses the confinement line determined from the disappearance of the infrared wall. For example, at $T=0.21~{\rm GeV}$ and $\omega=0.6~{\rm GeV}$, the variation with the chemical potential shown in the top-left panel crosses the thermodynamic first-order transition line and $E_c$ develops a clear jump. At the same temperature with $\omega=0$, on the other hand, the trajectory intersects only the confinement line extracted from the infrared-wall criterion, while $E_c$ remains completely smooth. The same behavior is observed in the remaining panels. This demonstrates that $E_c$ is primarily sensitive to the thermodynamic phase transition, whereas the disappearance of the infrared wall leaves no direct imprint on it.

\begin{figure}[h]
\begin{center}
\includegraphics[width=6.8cm]{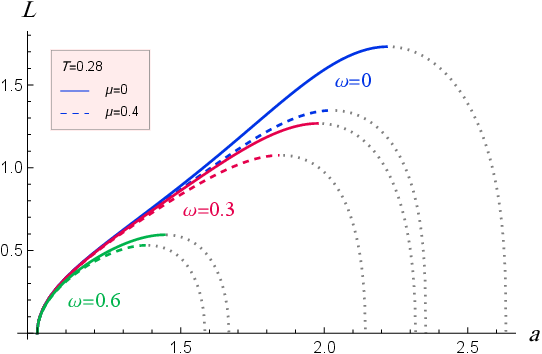}\hspace{.3cm}
\includegraphics[width=6.8cm]{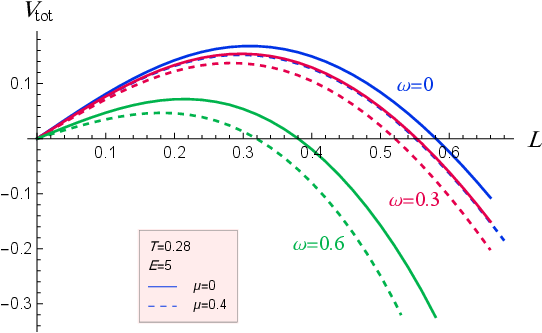}
\end{center}
\caption{\footnotesize 
Left: Separation length $L$ as a function of the rescaled turning point $a=z_c/z_0$ for $T=0.28~{\rm GeV}$, where all backgrounds are in the deconfined phase. Solid (dashed) colored curves correspond to the energetically favored connected-string solutions for $\mu=0$ ($\mu=0.4~{\rm GeV}$), while the dotted gray curves denote the second connected-string branch. Right: Total potential $V_{\rm tot}$ as a function of the quark--antiquark separation for the same backgrounds in the presence of a constant external electric field $E=5~{\rm GeV}^2$.}
\label{LVDeconfined}
\end{figure} 

The dependence of the separation length on the turning-point position is shown in Fig.\,\ref{LVDeconfined} for a representative deconfined background at $T=0.28~{\rm GeV}$. The left panel displays $L$ as a function of the rescaled turning point
\begin{align}
a=\frac{z_c}{z_0},
\end{align}
for several values of the angular velocity and chemical potential. The solid and dashed colored curves correspond to the energetically favored connected solutions for $\mu=0$ and $\mu=0.4~{\rm GeV}$, respectively, while the gray dotted curves represent the second connected branch with higher energy.

As in the static background, the separation length possesses a finite maximum value $L_{\rm max}$. For each separation smaller than $L_{\rm max}$ there exist two connected string solutions, but only the branch closer to the boundary is energetically preferred. Once the separation exceeds $L_{\rm max}$, no connected solution exists and the physical configuration is given by two disconnected straight strings extending toward the bulk.

Both the chemical potential and the angular velocity reduce the maximum separation length. At the same time, they move the turning point of the physical branch toward the boundary, thereby reducing the region of the bulk explored by the connected string. We shall return to this point later in this subsection, where its geometrical origin will be clarified by analyzing the radial domain accessible to the string.

\begin{figure}[h]
\begin{center}
\includegraphics[width=6.8cm]{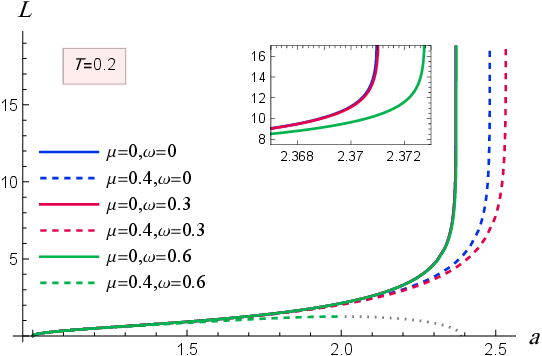}\hspace{.3cm}
\includegraphics[width=6.8cm]{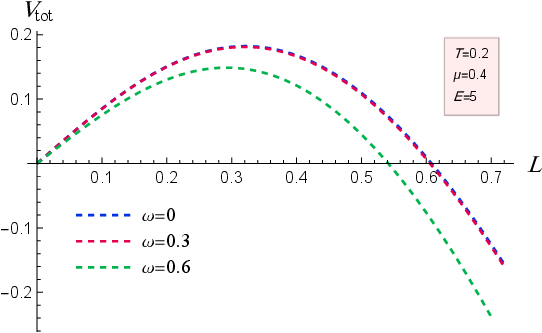}
\end{center}
\caption{\footnotesize 
Left: Separation length $L$ as a function of the rescaled turning point $a=z_c/z_0$ for backgrounds that are confined at vanishing angular velocity. The inset enlarges the $\mu=0$ curves to make the small effect of rotation visible. Right: Total potential $V_{\rm tot}$ for the $\mu=0.4~{\rm GeV}$ backgrounds under a constant external electric field $E=5~{\rm GeV}^2$. The case with $\omega=0.6~{\rm GeV}$ is in the deconfined phase according to the string criterion.}
\label{LVConfined}
\end{figure} 

The corresponding total potential is presented in the right panel of Fig.\,\ref{LVDeconfined} for an external electric field $E=5~{\rm GeV}^2$. Since the electric field is smaller than $E_c$ in all cases, as can be seen from Fig.\,\ref{Ecplots}, a finite potential barrier separates the vacuum from the pair-production state. Increasing either the chemical potential or the angular velocity lowers both the height and the width of this barrier, indicating that rotation facilitates the Schwinger pair-production process. Moreover, the influence of the angular velocity becomes progressively more significant at larger values of $\omega$, where the reduction of the potential barrier is considerably more pronounced.

The corresponding results for backgrounds that are confined at vanishing angular velocity are shown in Fig.\,\ref{LVConfined}. In the left panel, all the curves with $\mu=0$ and various angular velocities, together with the curves for $\mu=0.4~{\rm GeV}$ at $\omega=0$ and $0.3~{\rm GeV}$, exhibit the characteristic confining behavior: as the turning point approaches the infrared wall, the separation length diverges, reflecting the impossibility of separating the quark--antiquark pair by a finite distance. The case with $\mu=0.4~{\rm GeV}$ and $\omega=0.6~{\rm GeV}$, however, already lies in the deconfined phase according to the string criterion, as can be inferred from the $(T,\omega)$ phase diagram of Fig.\,\ref{phasediagramszw}. Consequently, its separation length displays the same finite maximum characteristic of the deconfined phase.

Although rotation always weakens confinement, its quantitative influence on the separation length is considerably milder in the confined phase than in the deconfined one. This is particularly evident at vanishing chemical potential, where the curves corresponding to different angular velocities almost overlap. To make these small differences visible, an enlarged view of the $\mu=0$ curves is shown in the inset. At nonzero chemical potential, the influence of rotation becomes more noticeable and increases with angular velocity, consistently with the behavior already observed for the infrared wall.

The right panel of Fig.\,\ref{LVConfined} presents the corresponding total potential for $\mu=0.4~{\rm GeV}$ under the same external electric field, whose magnitude lies within the interval $(E_s,E_c)$ for all the backgrounds considered. Rotation again reduces both the height and the width of the potential barrier, demonstrating that it enhances the Schwinger effect in the confined phase as well. Nevertheless, the reduction is significantly weaker than in the deconfined phase, in agreement with the comparatively small modifications induced by rotation in the infrared wall and the separation length. The corresponding potentials for $\mu=0$ are not shown because the changes produced by rotation are too small to be distinguished clearly on the scale of the figure and would obscure the more pronounced behavior observed at $\mu=0.4~{\rm GeV}$.

\begin{figure}[h]
\begin{center}
\includegraphics[width=7.4cm]{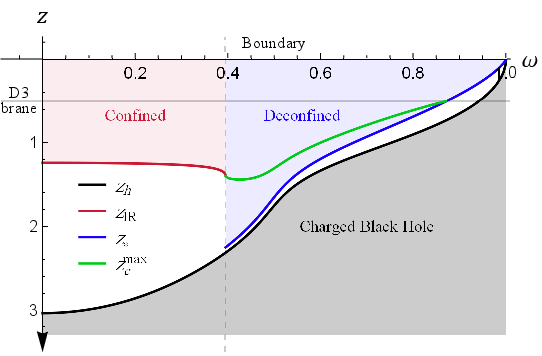}
\end{center}
\caption{\footnotesize 
Schematic evolution of the radial structure with the angular velocity for the representative background $(T,\mu)=(0.2,0.4)~{\rm GeV}$. The figure shows the event horizon $z_h$, the infrared wall $z_{\rm IR}$ in the confined phase, and the worldsheet horizon $z_*$ and the largest turning point of the energetically favored connected string solution, $z_c^{\rm max}$ in the deconfined phase. The shaded regions indicate the confined phase, the deconfined phase, and the black-hole interior. The horizontal line denotes the position of the probe D3-brane. For visualization purposes, the radial coordinate is plotted with the boundary at the top and increasing bulk depth toward the bottom.}
\label{geometry}
\end{figure} 

Finally, let us return to the geometrical interpretation of the maximum separation length discussed above. Figure\,\ref{geometry} summarizes the radial structure of the rotating background, for the representative background $(T,\mu)=(0.2,0.4)~{\rm GeV}$, by displaying the evolution of the event horizon, the infrared wall and the worldsheet horizon together with the largest radial position reached by the physical connected string solution with respect to $\omega$.

In the confined phase, the connected string can penetrate the bulk only up to the infrared wall located at $z_{\rm IR}$. As discussed in the previous subsection, increasing the angular velocity pushes this wall slightly deeper into the bulk, thereby enlarging the radial region accessible to the string and consequently weakening confinement. At the confinement--deconfinement transition, the infrared wall disappears abruptly. Simultaneously, the largest turning point of the energetically favored connected solution emerges continuously from the same radial position,
\begin{align}
z_c^{\rm max}(\omega_c)=z_{\rm IR}(\omega_c),
\end{align}
providing a smooth geometrical connection between the confined and deconfined descriptions.

Once the system enters the deconfined phase, the worldsheet horizon $z_*$ replaces the infrared wall as the upper limit of the region accessible to connected string configurations. Nevertheless, the physical connected string never reaches the worldsheet horizon itself. Instead, it is restricted to
\begin{align}
z_0\le z_c\le z_c^{\rm max}\le z_*\le z_h,
\end{align}
while the second, energetically disfavored branch extends toward $z_*$. As the angular velocity increases further, both $z_*$ and $z_c^{\rm max}$ move toward the boundary. Eventually they meet the probe D3-brane at the same critical angular velocity, beyond which no connected-string solution exists. In other words, the worldsheet horizon progressively compresses the region accessible to the connected string until it disappears completely.

\begin{figure}[h]
\begin{center}
\includegraphics[width=6.8cm]{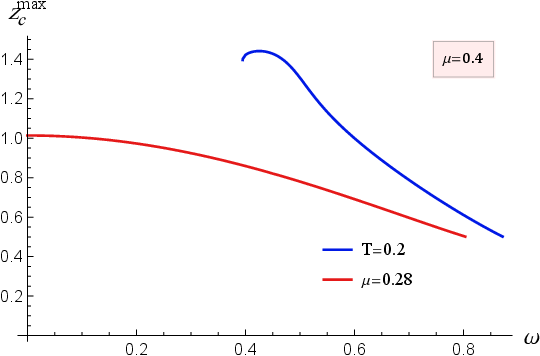}\hspace{.3cm}
\includegraphics[width=6.8cm]{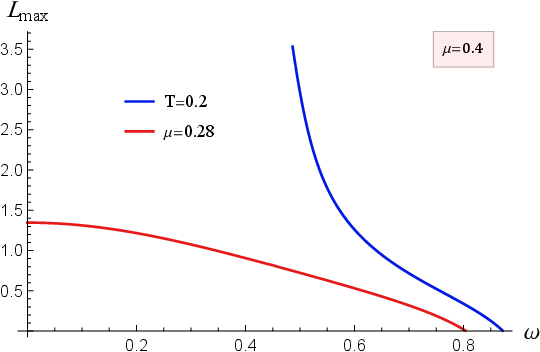}
\end{center}
\caption{\footnotesize 
Left: Largest turning point of the energetically favored connected string, $z_c^{\rm max}$, as a function of the angular velocity for two representative backgrounds. Right: Corresponding maximum separation length $L_{\rm max}$ as a function of the angular velocity for the same backgrounds. The probe D3-brane is located at $z_0=0.5\,{\rm GeV}^{-1}$.}
\label{Lmaxzcmax}
\end{figure} 

The behavior of $z_c^{\rm max}$ is examined quantitatively in the left panel of Fig.\,\ref{Lmaxzcmax}. For the background that is already deconfined at $\omega=0$, namely $(T,\mu)=(0.28,0.4)~{\rm GeV}$, the maximum turning point decreases monotonically with increasing angular velocity until it reaches the probe D3-brane at $z_0=0.5\,{\rm GeV}^{-1}$. In contrast, for the background $(T,\mu)=(0.2,0.4)~{\rm GeV}$, which is initially confined, $z_c^{\rm max}$ appears only after the confinement--deconfinement transition and initially follows the increasing motion of the disappearing infrared wall before eventually decreasing toward the probe brane. It should be emphasized that the precise form of $z_c^{\rm max}$ depends on the location of the probe D3-brane and therefore on the constituent quark mass.

The corresponding maximum separation length is shown in the right panel of Fig.\,\ref{Lmaxzcmax}. In both backgrounds, $L_{\rm max}$ decreases monotonically with the angular velocity, indicating that rotation reduces the largest distance over which a connected quark--antiquark configuration can remain stable. For the background that is deconfined at all angular velocities, this reduction is relatively gradual. By contrast, for the background that becomes deconfined only after crossing the phase transition, $L_{\rm max}$ decreases much more rapidly immediately after the transition before approaching zero at large angular velocity.

These observations provide a transparent geometrical explanation for the behavior of the total potential discussed above. By driving both the worldsheet horizon and the largest physical turning point toward the boundary, rotation forces the connected string to probe a progressively smaller portion of the bulk geometry. As a consequence, the maximum separation between the quark and antiquark decreases and the potential barrier against pair production becomes lower and narrower, thereby enhancing the Schwinger effect.

\section{Summary and discussion}
\label{sec:summary}

In this work, we investigated the effects of rotation on the phase structure and Schwinger pair production in a holographic QCD model based on the Einstein--Maxwell--dilaton framework. Rotation was introduced through the boost construction developed in \cite{huang}, which generates a stationary rotating background from the corresponding static solution while preserving the regularity of the geometry. This construction allowed us to study how angular velocity modifies both the thermodynamic properties of the plasma and the nonperturbative dynamics of quark--antiquark pairs in the presence of an external electric field.

We first analyzed the thermodynamic phase structure using the grand potential together with several independent response functions. The first-order transition line, the crossover trajectories extracted from the extrema of the thermodynamic observables, and the critical end point were found to obey a remarkably simple scaling law under rotation. All thermodynamic phase diagrams corresponding to different angular velocities collapse into the static one after the rescaling
\begin{align}
T\rightarrow\gamma T,
\qquad
\mu\rightarrow\gamma\mu,
\end{align}
demonstrating that the entire thermodynamic sector is governed by a universal boost relation. As a consequence, the influence of rotation on the thermodynamic observables is completely determined by the corresponding static background.

We then investigated confinement from the viewpoint of the string degrees of freedom through the effective string tension. Unlike the thermodynamic observables, the confinement criterion determined by the existence of the dynamical infrared wall does not satisfy the same scaling relation. Consequently, although rotation again shifts the confinement--deconfinement transition toward lower temperatures and chemical potentials, the corresponding phase boundaries are no longer related to those of the static theory by the universal boost transformation governing the thermodynamic sector. 
As in the static EMD model studied in \cite{superus}, this confinement line separates from the thermodynamic first-order transition line before reaching the critical end point and extends into the supercritical region. We found here that rotation enhances this separation, thereby enlarging the region where thermodynamic observables still characterize the system as confined while the quark potential already exhibits deconfined behavior. These results demonstrate that, although both sectors respond qualitatively similarly to rotation, the string sector acquires genuinely nontrivial rotational modifications beyond those dictated by the boost transformation, revealing complementary aspects of the underlying strongly coupled dynamics.

The potential analysis revealed several new features induced by rotation. In the confined phase, rotation pushes the infrared wall slightly deeper into the bulk, reducing the confining string tension and consequently lowering the critical electric field $E_s$. In the deconfined phase, however, rotation produces a much more significant modification through the appearance of a worldsheet horizon. Unlike the static case, where the connected string solutions may extend continuously toward the black-hole horizon, the rotating geometry restricts them to a finite radial interval ending at the worldsheet horizon. As the angular velocity increases, this worldsheet horizon moves monotonically toward the boundary, progressively reducing the portion of the bulk accessible to the connected string.

This geometric restriction has direct physical consequences. Both the maximum separation length of the quark--antiquark pair and the deepest turning point reached by the energetically favored connected string decrease with increasing angular velocity. The total potential barrier responsible for vacuum tunneling is simultaneously reduced in both height and width. Rotation therefore enhances the Schwinger effect in both confined and deconfined phases. The reduction is particularly pronounced in the deconfined phase, where the worldsheet horizon strongly limits the string dynamics, while the modifications remain comparatively mild in the confined phase because the infrared wall itself is only weakly affected by rotation.

The critical electric field $E_c$, determined by the probe D3-brane, was also found to decrease monotonically with increasing angular velocity. Since the observation of the Schwinger effect in heavy-ion collisions is largely hindered by the extremely large electric fields required for pair production, the reduction of both critical fields by rotation provides a favorable condition for vacuum pair production. Interestingly, while the confining critical field $E_s$ is entirely controlled by the disappearance of the infrared wall, the critical field $E_c$ is primarily sensitive to the thermodynamic first-order transition and exhibits discontinuous jumps only when this transition is crossed. This illustrates that the two critical fields are governed by different geometric mechanisms in the holographic description.

The present work naturally suggests several directions for future investigation. One immediate extension would be to study Schwinger pair production in rotating holographic backgrounds constructed through alternative realizations of rotation, such as those obtained by transforming the static geometry to cylindrical coordinates followed by an angular transformation. Such constructions permit independent investigations of quark--antiquark pairs oriented both parallel and perpendicular to the rotation axis, providing direct access to rotational anisotropies in the Schwinger effect.
Such constructions would also make it possible to disentangle the anisotropy induced by rotation from anisotropies generated by other external sources, such as strong magnetic fields, allowing a systematic comparison of their effects on the Schwinger mechanism, the effective string tension, and the confinement properties of quark--antiquark pairs along different spatial directions.
Another interesting direction would be to investigate Schwinger pair production in the recently constructed fully backreacted anisotropic rotating Einstein--Maxwell--dilaton backgrounds, which successfully reproduce the enhancement of the transition temperature observed in present lattice simulations. Comparing the resulting string observables with those obtained in the boost construction employed here could help disentangle model-dependent features from more universal properties of rotating strongly coupled plasmas.
 It would also be interesting to explore whether the worldsheet horizon identified here admits a more general interpretation within rotating holographic plasmas and to investigate its relation to other nonequilibrium worldsheet horizons encountered in holographic transport phenomena. Finally, extending the present analysis to time-dependent electromagnetic fields or to backgrounds including magnetic fields would provide a more realistic description of the environment created in relativistic heavy-ion collisions.



\begin{thebibliography}{9}

\bibitem{Bmagnitude1}
V. Skokov, A. Yu. Illarionov, and V. Toneev, ``Estimate of the magnetic field strength in heavy-ion collisions,'' Int. J. Mod. Phys. A {\bf 24}, 5925 (2009) \href{https://arxiv.org/abs/0907.1396}{[arXiv:0907.1396 [nucl-th]]}. 

\bibitem{Bmagnitude2}
W.-T. Deng and X.-G. Huang, ``Event-by-event generation of electromagnetic fields in heavy-ion collisions,'' Phys. Rev. C {\bf 85}, 044907 (2012) \href{https://arxiv.org/abs/1201.5108}{[arXiv:1201.5108 [nucl-th]]}. 

\bibitem{angularm}
F. Becattini, F. Piccinini, and J. Rizzo, ``Angular momentum conservation in heavy ion collisions at very high energy,'' Phys. Rev. C {\bf 77}, 024906 (2008) \href{https://arxiv.org/abs/0711.1253}{[arXiv:0711.1253 [nucl-th]]}. 

\bibitem{star2017}
STAR Collaboration, ``Global $\Lambda$ hyperon polarization in nuclear collisions: evidence for the most vortical fluid,'' Nature {\bf 548}, 62 (2017) \href{https://arxiv.org/abs/1701.06657}{[arXiv:1701.06657 [nucl-ex]]}. 

\bibitem{hydro}
Y. Jiang, Z.-W. Lin, and J. Liao, ``Rotating quark-gluon plasma in relativistic heavy ion collisions,'' Phys. Rev. C {\bf 94}, 044910  (2016) \href{https://arxiv.org/abs/1602.06580}{[arXiv:1602.06580 [hep-ph]]}. 

\bibitem{alice}
ALICE Collaboration, ``Evidence of spin-orbital angular momentum interactions in relativistic heavy-ion collisions,'' Phys. Rev. Lett. {\bf 125}, 012301 (2020). 

\bibitem{star2023}
STAR Collaboration, ``Pattern of Global Spin Alignment of $\phi$ and $K^{*0}$ mesons in Heavy-Ion Collisions,'' Nature {\bf 614}, 7947, 244-248 (2023) \href{https://arxiv.org/abs/2204.02302}{[arXiv:2204.02302 [hep-ph]]}. 

\bibitem{chiralvortical1}
O. Rogachevsky, A. Sorin, and O. Teryaev, ``Chiral vortaic effect and neutron asymmetries in heavy-ion collisions,'' Phys. Rev. C {\bf 82}, 054910 (2010) \href{https://arxiv.org/abs/1006.1331}{[arXiv:1006.1331 [hep-ph]]}. 

\bibitem{chiralvortical2}
D. E. Kharzeev, J. Liao, S. A. Voloshin, and G. Wang, ``Chiral magnetic and vortical effects in high-energy nuclear collisions --- A status report,'' Prog. Part. Nucl. Phys. {\bf 88}, 1 (2016) \href{https://arxiv.org/abs/1511.04050}{[arXiv:1511.04050 [hep-ph]]}. 

\bibitem{chiralvortical3}
S.-Y. Yang, R.-H. Fang, D.-F. Hou, and H.-C. Ren, ``Axial chiral vortical effect in a sphere with finite size effect*,'' Chin. Phys. C {\bf 47}, 034106 (2023) \href{https://arxiv.org/abs/2111.13053}{[arXiv:2111.13053 [hep-th]]}. 

\bibitem{polarization1}
Y. Guo, J. Liao, E. Wang, H. Xing, and H. Zhang, ``Hyperon polarization from the vortical fluid in low-energy nuclear collisions,'' Phys. Rev. C {\bf 104}, L041902 (2021) \href{https://arxiv.org/abs/2105.13481}{[arXiv:2105.13481 [nucl-th]]}. 

\bibitem{polarization2}
H. A. Ahmed, Y. Chen, and M. Huang, ``Gluon polarization contribution to the spin alignment of vector mesons from holography,'' Phys. Rev. D {\bf 111}, L086006 (2025) \href{https://arxiv.org/abs/2501.13401}{[arXiv:2501.13401 [hep-ph]]}. 

\bibitem{polarization3}
V. V. Braguta, M. N. Chernodub, and A. A. Roenko, ``Chiral and deconfinement thermal transitions at finite quark spin polarization in lattice QCD simulations,'' Phys. Rev. D {\bf 111}, 114508  (2025) \href{https://arxiv.org/abs/2503.18636}{[arXiv:2503.18636 [hep-lat]]}. 

\bibitem{transition1}
M. N. Chernodub and S. Gongyo, ``Interacting fermions in rotation: chiral symmetry restoration, moment of inertia and thermodynamics,'' JHEP {\bf 01}, 136 (2017) \href{https://arxiv.org/abs/1611.02598}{[arXiv:1611.02598 [hep-th]]}. 

\bibitem{huang}
X. Chen, L. Zhang, D. Li, D. Hou, and M. Huang,  ``Gluodynamics and deconfinement phase transition under rotation from holography,'' JHEP {\bf 07}, 132 (2021) \href{https://arxiv.org/abs/2010.14478}{[arXiv:2010.14478 [hep-ph]]}.

\bibitem{transition2}
M. N. Chernodub, ``Inhomogeneous confining-deconfining phases in rotating plasmas,'' Phys. Rev. D {\bf 103}, 054027  (2021) \href{https://arxiv.org/abs/2012.04924}{[arXiv:2012.04924 [hep-ph]]}. 

\bibitem{transition3}
N. R.F. Braga, L. F. Faulhaber, and O. C. Junqueira, ``Confinement-deconfinement temperature for a rotating quark-gluon plasma,'' Phys. Rev. D {\bf 105}, 106003  (2022) \href{https://arxiv.org/abs/2201.05581}{[arXiv:2201.05581 [hep-th]]}. 

\bibitem{transition4}
J.-H. Wang and S.-Q. Feng, ``Rotation effect on the deconfinement phase transition in holographic QCD,'' Phys. Rev. D {\bf 109}, 066019  (2024) \href{https://arxiv.org/abs/2403.01814}{[arXiv:2403.01814 [hep-ph]]}. 

\bibitem{transition5}
N. R. F. Braga and A. L. Ferreira Jr, ``Unraveling the effect of rotation on the confinement/deconfinement transition of the quark-gluon plasma,''  \href{https://arxiv.org/abs/2511.22464}{[arXiv:2511.22464 [hep-th]]}. 

\bibitem{effective1}
H.-L. Chen, K. Fukushima, X.-G. Huang, and K. Mameda, ``Analogy between rotation and density for Dirac fermions in a magnetic field,'' Phys. Rev. D {\bf 93}, 104052 (2016) \href{https://arxiv.org/abs/1512.08974}{[arXiv:1512.08974 [hep-ph]]}. 

\bibitem{effective2}
Y. Jiang and J. Liao, ``Pairing phase transitions of matter under rotation,'' Phys. Rev. Lett. {\bf 117}, 192302 (2016) \href{https://arxiv.org/abs/1606.03808}{[arXiv:1606.03808 [hep-ph]]}. 

\bibitem{effective3}
M. N. Chernodub and S. Gongyo, ``Effects of rotation and boundaries on chiral symmetry breaking of relativistic fermions,'' Phys. Rev. D. {\bf 95}, 096006 (2017) \href{https://arxiv.org/abs/1702.08266}{[arXiv:1702.08266 [hep-th]]}. 

\bibitem{effective4}
S. Ebihara, K. Fukushima, and K. Mameda, ``Boundary effects and gapped dispersion in rotating fermionic matter,'' Phys. Lett. B {\bf 764}, 94 (2017) \href{https://arxiv.org/abs/1608.00336}{[arXiv:1608.00336 [hep-ph]]}. 

\bibitem{effective6}
X. Wang, M. Wei, Z. Li, and M. Huang, ``Quark matter under rotation in the NJL model with vector interaction,'' Phys. Rev. D {\bf 99}, 016018 (2019) \href{https://arxiv.org/abs/1808.01931}{[arXiv:1808.01931 [hep-ph]]}. 

\bibitem{effective7}
Y. Fujimoto, K. Fukushima, and Y. Hidaka, ``Deconfining phase boundary of rapidly rotating hot and dense matter and analysis of moment of inertia,'' Phys. Lett. B {\bf 816}, 136184 (2021) \href{https://arxiv.org/abs/2101.09173}{[arXiv:2101.09173 [hep-ph]]}. 

\bibitem{effective8}
F. Sun and A. Huang, ``Properties of strange quark matter under strong rotation,'' Phys. Rev. D {\bf 106}, 076007 (2022) \href{https://arxiv.org/abs/2104.14382}{[arXiv:2104.14382 [hep-ph]]}. 

\bibitem{effective9}
K. Xu, F. Lin, A. Huang, and M. Huang, ``$\Lambda/{\bar \Lambda}$ polarization and splitting induced by rotation and magnetic field,'' Phys. Rev. D {\bf 106}, L071502 (2022) \href{https://arxiv.org/abs/2205.02420}{[arXiv:2205.02420 [hep-ph]]}. 

\bibitem{effective10}
F. Sun, K. Xu, and M. Huang, ``Splitting of chiral and deconfinement phase transitions induced by rotation,'' Phys. Rev. D {\bf 108}, 096007 (2023) \href{https://arxiv.org/abs/2307.14402}{[arXiv:2307.14402 [hep-ph]]}. 

\bibitem{lattice1}
V. V. Braguta, A.Yu. Kotov, D. D. Kuznedelev, and A. A. Roenko, ``Influence of relativistic rotation on the confinement-deconfinement transition in gluodynamics,'' Phys. Rev. D {\bf 103}, 094515 (2021) \href{https://arxiv.org/abs/2102.05084}{[arXiv:2102.05084 [hep-lat]]}. 

\bibitem{lattice2}
V. V. Braguta, A. Yu. Kotov, A. A. Roenko, and D. A. Sychev, ``Thermal phase transitions in rotating QCD with dynamical quarks,'' {\it PoS} LATTICE{\bf 2022}, 190 (2023) \href{https://arxiv.org/abs/2212.03224}{[arXiv:2212.03224 [hep-lat]]}. 

\bibitem{lattice3}
J.-C. Yang and X.-G. Huang, ``QCD on rotating lattice with staggered fermions,'' \href{https://arxiv.org/abs/2307.05755}{[arXiv:2307.05755 [hep-lat]]}. 

\bibitem{correctphase1}
F. Sun, J. Shao, R. Wen, K. Xu, and M. Huang, ``Chiral phase transition and spin alignment of vector meson in the Polarized-Polyakov-loop Nambu-Jona-Lasinio model under rotation,'' Phys. Rev. D {\bf 109}, 116017 (2024) \href{https://arxiv.org/abs/2402.16595}{[arXiv:2402.16595 [hep-ph]]}. 

\bibitem{correctphase2}
Y. Chen, X. Chen, D. Li, and M. Huang, ``Deconfinement and chiral restoration phase transition under rotation from holography in an anisotropic gravitational background,'' Phys. Rev. D {\bf 111}, 046006 (2025) \href{https://arxiv.org/abs/2405.06386}{[arXiv:2405.06386 [hep-ph]]}. 

\bibitem{adscft1}
  S. S. Gubser, I. R. Klebanov, and A. M. Polyakov, ``Gauge theory correlators from non-critical string theory,''   Phys. Lett.  B {\bf 428}, 105 (1998) \href{http://arxiv.org/abs/hep-th/9802109}{[arXiv:hep-th/9802109]}.

\bibitem{adscft2}
  E. Witten, ``Anti-de Sitter space and holography,'' Adv. Theor. Math. Phys.  {\bf 2}, 253 (1998) \href{http://arxiv.org/abs/hep-th/9802150}{[arXiv:hep-th/9802150]}.

\bibitem{adscft3}
  J. M. Maldacena, ``The large N limit of superconformal field theories and supergravity,'' Adv. Theor. Math. Phys.  {\bf 2}, 231 (1998)
  [Int. J. Theor. Phys.  {\bf 38}, 1113 (1999)]  \href{http://arxiv.org/abs/hep-th/9711200}{[arXiv:hep-th/9711200]}.

\bibitem{adscft4} 
  J. Casalderrey-Solana, H. Liu, D. Mateos, K. Rajagopal, and U. A. Wiedemann, ``Gauge/string duality, hot QCD and heavy ion collisions,'' Cambridge, UK: Cambridge University Press, 2014 \href{http://arxiv.org/abs/1101.0618}{[arXiv:1101.0618 [hep-th]]}.

\bibitem{KerrAds1} 
S. W. Hawking, C. J. Hunter, and M. M. Taylor-Robinson, ``Rotation and the AdS / CFT correspondence,'' Phys. Rev. D {\bf 59}, 064005 (1999) \href{https://arxiv.org/abs/hep-th/9811056}{[arXiv:hep-th/9811056]}.

\bibitem{KerrAds2} 
B. McInnes, ``Angular momentum in QGP holography,'' Nucl. Phys. B{\bf 887}, 246 (2014) \href{https://arxiv.org/abs/1403.3258}{[arXiv:1403.3258 [hep-th]]}.

\bibitem{KerrAds3} 
C. Erices and C. Martinez, ``Rotating hairy black holes in arbitrary dimensions,'' Phys. Rev. D {\bf 97}, 024034 (2018) \href{https://arxiv.org/abs/1707.03483}{[arXiv:1707.03483 [hep-th]]}.

\bibitem{KerrAds4} 
B. McInnes, ``Viscosity vs. Vorticity in the Quark-Gluon Plasma,'' Nucl. Phys. B{\bf 953}, 114951 (2020) \href{https://arxiv.org/abs/1812.07146}{[arXiv:1812.07146 [hep-th]]}.

\bibitem{KerrAds5} 
I. Y. Aref'eva, A. A. Golubtsova, and E. Gourgoulhon, ``Holographic drag force in 5d Kerr-AdS black hole,'' JHEP {\bf 04}, 169 (2021) \href{https://arxiv.org/abs/2004.12984}{[arXiv:2004.12984 [hep-th]]}.

\bibitem{KerrAds6} 
A. A. Golubtsova and N. S. Tsegelnik, ``Probing the holographic model of ${\cal N}=4$ SYM rotating quark-gluon plasma,'' Phys. Rev. D {\bf 107}, 106017 (2023) \href{https://arxiv.org/abs/2211.11722}{[arXiv:2211.11722 [hep-th]]}.

\bibitem{boost1} 
A. Awad, ``Higher dimensional charged rotating solutions in (A)dS space-times,''  Class.Quant.Grav. {\bf 20}, 2827 (2003) \href{https://arxiv.org/abs/hep-th/0209238}{[arXiv:hep-th/0209238]}.

\bibitem{boost2} 
A. Sheykhi and S.H. Hendi, ``Charged rotating dilaton black branes in AdS universe,'' Gen.Rel.Grav. {\bf 42}, 1571 (2010) \href{https://arxiv.org/abs/0911.2831}{[arXiv:0911.2831 [hep-th]]}.

\bibitem{boost3} 
M. B. Gaete, L. Guajardo, and M. Hassaine, ``A Cardy-like formula for rotating black holes with planar horizon,'' JHEP {\bf 04}, 092 (2017) \href{https://arxiv.org/abs/1702.02416}{[arXiv:1702.02416 [hep-th]]}.

\bibitem{boost4} 
Y.-Q. Zhao, S. He, D. Hou, L. Li, and Z. Li, ``Phase diagram of holographic thermal dense QCD matter with rotation,'' JHEP {\bf 04}, 115 (2023) \href{https://arxiv.org/abs/2212.14662}{[arXiv:2212.14662 [hep-ph]]}.

\bibitem{boost5} 
 J. Zhou, X. Chen, Y.-Q. Zhao, and J. Ping, ``Thermodynamics of heavy quarkonium in rotating matter from holography,'' Phys. Rev. D {\bf 102}, 126029  (2020).

\bibitem{boost6} 
 Z.-q. Zhang, X. Zhu, and D.-f. Hou, ``Imaginary potential of heavy quarkonia from thermal fluctuations in rotating matter from holography,'' Nucl. Phys. B{\bf 989}, 116149 (2023).


\bibitem{planar} 
J.-X. Chen, D.-F. Hou, and H.-C. Ren, ``Drag force and heavy quark potential in a rotating background,'' JHEP {\bf 03}, 171 (2024) \href{https://arxiv.org/abs/2308.08126}{[arXiv:2308.08126 [hep-ph]]}.


\bibitem{rotatinggauge1} 
O. Domenech, M. Montull, A. Pomarol, A. Salvio, and P. J. Silva, ``Emergent gauge fields in holographic superconductors,'' JHEP {\bf 08}, 033 (2010) \href{https://arxiv.org/abs/1005.1776}{[arXiv:1005.1776 [hep-th]]}.

\bibitem{rotatinggauge2} 
V. Keranen, E. Keski-Vakkuri, S. Nowling, and K. P. Yogendran, ``Inhomogeneous Structures in Holographic Superfluids: II. Vortices,'' Phys. Rev. D {\bf 81}, 126012 (2010) \href{https://arxiv.org/abs/0912.4280}{[arXiv:0912.4280 [hep-th]]}.

\bibitem{rotatinggauge3} 
O. J. C. Dias, G. T. Horowitz, N. Iqbal, and J. E. Santos, ``Vortices in holographic superfluids and superconductors as conformal defects,'' JHEP {\bf 04}, 096 (2014) \href{https://arxiv.org/abs/1311.3673}{[arXiv:1311.3673 [hep-th]]}.

\bibitem{rotatinggauge4} 
Y. Chen, D. Li, and M. Huang, ``Inhomogeneous chiral condensation under rotation in the holographic QCD,'' Phys. Rev. D {\bf 106}, 106002 (2022) \href{https://arxiv.org/abs/2208.05668}{[arXiv:2208.05668 [hep-ph]]}.

\bibitem{dudalm}
D. Dudal and S. Mahapatra, ``Thermal entropy of a quark-antiquark pair above and below deconfinement from a dynamical holographic QCD model,'' Phys. Rev. D {\bf 96}, 126010 (2017) \href{https://arxiv.org/abs/1708.06995}{[arXiv:1708.06995 [hep-th]]}. 

\bibitem{Schwinger}
  J. S. Schwinger,
  ``On gauge invariance and vacuum polarization,''
  Phys. Rev. {\bf 82}, 664 (1951).

\bibitem{semenoff}
G. W. Semenoff and K. Zarembo, ``Holographic Schwinger effect,''
Phys. Rev. Lett. {\bf 107}, 171601 (2011) \href{https://arxiv.org/abs/1109.2920}{[arXiv:1109.2920 [hep-th]]}.

\bibitem{potential}
Y. Sato and K. Yoshida, ``Potential analysis in holographic Schwinger effect,'' JHEP {\bf 08}, 002 (2013) \href{https://arxiv.org/abs/1304.7917}{[arXiv:1304.7917 [hep-th]]}.

\bibitem{Sch1}
Y. Sato and K. Yoshida, ``Universal aspects of holographic Schwinger effect in general backgrounds,'' JHEP {\bf 12}, 051 (2013) \href{https://arxiv.org/abs/1309.4629}{[arXiv:1309.4629 [hep-th]]}.

\bibitem{Sch2}
Y. Sato and K. Yoshida, ``Holographic description of the Schwinger effect in electric
and magnetic fields,'' JHEP {\bf 04}, 111 (2013) \href{https://arxiv.org/abs/1303.0112}{[arXiv:1303.0112 [hep-th]]}.

\bibitem{Sch3}
S. Bolognesi, F. Kiefer, and E. Rabinovici, ``Comments on critical electric and magnetic fields from holography,'' JHEP {\bf 01}, 174 (2013) \href{https://arxiv.org/abs/1210.4170}{[arXiv:1210.4170 [hep-th]]}.

\bibitem{confin1}
Y. Sato and K. Yoshida, ``Holographic Schwinger effect in confining phase,'' JHEP {\bf 09}, 134 (2013) \href{https://arxiv.org/abs/1306.5512}{[arXiv:1306.5512 [hep-th]]}.

\bibitem{confin2}
D. Kawai, Y. Sato, and K. Yoshida, ``The Schwinger pair production rate in confining
theories via holography,'' Phys. Rev. D {\bf 89}, 101901 (2014) \href{https://arxiv.org/abs/1312.4341}{[arXiv:1312.4341 [hep-th]]}.

\bibitem{Sch4}
M. Ghodrati, ``Schwinger effect and entanglement entropy in confining geometries,'' Phys. Rev. D {\bf 92}, 065015 (2015) \href{https://arxiv.org/abs/1506.08557}{[ 	arXiv:1506.08557 [hep-th]]}.

\bibitem{Sch5}
S-J. Zhang and E. Abdalla, ``Holographic Schwinger effect in a confining background with Gauss-Bonnet corrections,'' Gen. Rel. Grav. {\bf 48}, 60 (2016) \href{https://arxiv.org/abs/1508.03364}{[arXiv:1508.03364 [hep-th]]}.

\bibitem{confinrev}
D. Kawai, Y. Sato, and K. Yoshida, ``A holographic description of the Schwinger effect in a confining gauge theory,'' Int. J. Mod. Phys. A {\bf 30}, 1530026 (2015) \href{https://arxiv.org/abs/1504.00459}{[arXiv:1504.00459 [hep-th]]}.

\bibitem{dehghani}
L. Shahkarami, M. Dehghani, and P. Dehghani, ``Holographic Schwinger effect in a D-instanton background,'' Phys. Rev. D {\bf 97}, 046013 (2018)
 \href{https://arxiv.org/abs/1511.07986}{[arXiv:1511.07986 [hep-th]]}.

\bibitem{magneticdecay}
K. Hashimoto, T. Oka, and A. Sonoda, ``Electromagnetic instability in holographic QCD,'' JHEP {\bf 06}, 001 (2015) \href{https://arxiv.org/abs/1412.4254}{[arXiv:1412.4254 [hep-th]]}.

\bibitem{LF}
L. Shahkarami and F. Charmchi, ``Confining D-instanton background in an external electric field,'' Eur. Phys. J. C {\bf 79}, 343 (2019) \href{https://arxiv.org/abs/1904.09806}{[arXiv:1904.09806 [hep-th]]}.

 \bibitem{us2}
D. Masoumi, L. Shahkarami, and F. Charmchi, ``Effect of electromagnetic fields on deformed $\mathrm{AdS}_5$ models,'' Phys. Rev. D {\bf 101}, 126011 (2020) \href{https://arxiv.org/abs/2003.06848}{[arXiv:2003.06848 [hep-th]]}.

\bibitem{me}
L. Shahkarami, ``Magnetized Einstein-Maxwell-dilaton model under external electric fields,'' Eur. Phys. J. C {\bf 82}, 33 (2022)
 \href{https://arxiv.org/abs/2111.04813}{[arXiv:2111.04813 [hep-th]]}. 

 \bibitem{Sch6}
Sw. Li, Sk. Luo, and Hq. Li, ``Holographic Schwinger effect and electric instability with anisotropy,'' JHEP {\bf 08}, 206 (2022)
 \href{https://arxiv.org/abs/2205.01885}{[arXiv:2205.01885 [hep-th]]}.

\bibitem{Sch7}
S. Lin, X. Liu, X. Chen, G.-F. Zhang, and J. Zhou, ``Holographic Schwinger effect in flavor-dependent systems,'' Phys. Rev. D {\bf 111}, 046005 (2025) \href{https://arxiv.org/abs/2407.14828}{[arXiv:2407.14828 [hep-ph]]}.

\bibitem{Sch8}
R. Zhou and Z.-R. Zhu, ``Holographic Schwinger effect in strongly coupled ${\cal N}=4$ super Yang–Mills plasma on the Coulomb branch,'' Eur. Phys. J. C {\bf 85}, 046005 (2025).

\bibitem{rotSch1}
Y.-z. Cai, R.-p. Jing, and Z.-q. Zhang, ``Holographic Schwinger effect in a rotating strongly coupled medium,'' Chin. Phys. C {\bf 46}, 104107 (2022) \href{https://arxiv.org/abs/2206.15052}{[arXiv:2206.15052 [hep-th]]}.

\bibitem{rotSch2}
H. Xu, M. Ilyas, and Y.-C. Huang, ``Holographic Schwinger effect with a rotating probe D3-brane,'' Adv. High Energy Phys. {\bf 2023}, 6614276 (2023) \href{https://arxiv.org/abs/1604.06331}{[arXiv:1604.06331 [hep-th]]}.

\bibitem{rotSch3}
Y.-z. Cai and Z.-q. Zhang, ``Holographic Schwinger effect in spinning black hole backgrounds*,'' Chin. Phys. C {\bf 48}, 015102 (2024) \href{https://arxiv.org/abs/2310.13865}{[arXiv:2310.13865 [hep-ph]]}.

\bibitem{superus}
L. Shahkarami and F. Charmchi, ``Schwinger effect in dynamical holographic QCD with a supercritical region,'' JHEP {\bf 02}, 014 (2026) \href{https://arxiv.org/abs/2509.20151}{[arXiv:2509.20151 [hep-th]]}.

\bibitem{reconst1}
J. Alanen, K. Kajantie, and V. Suur-Uski, ``Gauge/gravity duality model for gauge theory thermodynamics,'' Phys. Rev. D {\bf 80}, 126008 (2009) \href{https://arxiv.org/abs/0911.2114}{[arXiv:0911.2114 [hep-ph]]}.

\bibitem{reconst2}
D. Li, S. He, M. Huang, and Q. S. Yan, ``Thermodynamics of deformed AdS$_5$ model with a positive/negative quadratic correction in graviton-dilaton system,'' JHEP {\bf 09}, 041 (2011) \href{https://arxiv.org/abs/1103.5389}{[arXiv:1103.5389 [hep-th]]}.

\bibitem{reconst3}
S. He, S. Y. Wu, Y. Yang, and P. H. Yuan, ``Phase structure in a dynamical soft-wall holographic QCD model,'' JHEP {\bf 04}, 093 (2013) \href{https://arxiv.org/abs/1301.0385}{[arXiv:1301.0385 [hep-th]]}.

\bibitem{reconst4}
Y. Yang and P.-H. Yuan, ``Confinement-deconfinement phase transition for heavy quarks in a soft wall holographic QCD model,'' JHEP {\bf 12}, 161 (2015) \href{https://arxiv.org/abs/1506.05930}{[arXiv:1506.05930 [hep-th]]}.

\bibitem{reconst5}
I. Aref'eva and K. Rannu, ``Holographic anisotropic background with confinement-deconfinement phase transition,'' JHEP {\bf 05}, 206 (2018) \href{https://arxiv.org/abs/1802.05652}{[arXiv:1802.05652 [hep-th]]}.    

\bibitem{hajilou}
H. Bohra, D. Dudal, A. Hajilou, and S. Mahapatra, ``Anisotropic string tensions and inversely magnetic catalyzed deconfinement from a dynamical AdS/QCD model,'' Phys. Lett. B {\bf 801}, 135184 (2020) \href{https://arxiv.org/abs/1907.01852}{[arXiv:1907.01852 [hep-th]]}. 

\bibitem{zstar1}
S. S. Gubser, ``Drag force in AdS/CFT,'' Phys. Rev. D {\bf 74}, 126005 (2006) \href{https://arxiv.org/abs/hep-th/0605182}{[arXiv:hep-th/0605182]}.    

\bibitem{zstar2}
C. P. Herzog, A. Karch, P. Kovtun, C. Kozcaz, and L. G. Yaffe, ``Energy loss of a heavy quark moving through ${\cal N}=4$ supersymmetric Yang-Mills plasma,'' JHEP  {\bf 07}, 013 (2006) \href{https://arxiv.org/abs/hep-th/0605158}{[arXiv:hep-th/0605158]}.  

 \bibitem{zstar3}
M. Chernicoff, J. A. Garcia, and A. Guijosa, ``The energy of a moving quark-antiquark pair in an ${\cal N}=4$ SYM plasma,'' JHEP  {\bf 09}, 068 (2006) \href{https://arxiv.org/abs/hep-th/0607089}{[arXiv:hep-th/0607089]}. 

 \bibitem{zstar4}
E. Caceres, M. Natsuume, and T. Okamura, ``Screening length in plasma winds,'' JHEP  {\bf 10}, 011 (2006) \href{https://arxiv.org/abs/hep-th/0607233}{[arXiv:hep-th/0607233]}.  

\bibitem{zstar5}
H. Liu, K. Rajagopal, and U. A. Wiedemann, ``An AdS/CFT Calculation of Screening in a Hot Wind,'' Phys. Rev. Lett.  {\bf 98}, 182301 (2007) \href{https://arxiv.org/abs/hep-ph/0607062}{[arXiv:hep-ph/0607062]}.    

\bibitem{zstar6}
A. N. Atmaja and K. Schalm, ``Anisotropic Drag Force from 4D Kerr-AdS Black Holes,'' JHEP  {\bf 04}, 070 (2011) \href{https://arxiv.org/abs/1012.3800}{[arXiv:1012.3800 [hep-th]]}.    


\end{thebibliography}
 \end{document}